\documentclass[3p,sort&compress,letterpaper,12pt,times,number]{elsarticle}

\usepackage{xspace}
\usepackage{amsmath}
\usepackage{amssymb}
\usepackage{color}

\usepackage{graphicx}
\usepackage{epstopdf}
\usepackage[bookmarks%
  ,pdftitle={An artificial neural network based $b$ jet identification algorithm at the CDF Experiment}%
  ,pdfsubject={CDF b tagging algorithm}%
  ,pdfauthor={ketchum, lewis, poprocki, freeman, pronko, rusu, wittich}%
  ,pdfstartview={Fit}%
  ]{hyperref}
\usepackage[all]{hypcap}
\newcommand{\mett}{\mbox{$\not\!\!E_{T}$}\xspace}
\newcommand{\ppbar}{\ensuremath{p\overline{p}}\xspace}

\graphicspath{{figures_nim/},{figures/}, {epsfig/}}

\journal{Nuclear Instruments and Methods A}

\begin{document}

\begin{frontmatter}


\title{\hfill{\small CDF/PUB/SEC\_VTX/PUBLIC/10631}\\[-11pt]
\hfill{\small FERMILAB-TM-2515-E-PPD}\\[5pt]
An artificial neural network based $b$ jet identification algorithm at the CDF Experiment}

\author[fnal]{J.~Freeman}
\author[chicago]{W.~Ketchum}
\author[fnal]{J.D.~Lewis}
\author[cornell]{S.~Poprocki}
\author[fnal]{A.~Pronko\fnref{fn1}}
\author[fnal]{V.~Rusu}
\author[cornell]{P.~Wittich\corref{cor1}}
\ead{wittich@cornell.edu}

\cortext[cor1]{Corresponding author}
\address[fnal]{Fermi National Accelerator Laboratory, Batavia, IL 60510}
\address[chicago]{University of Chicago, Chicago, IL, 60637}
\address[cornell]{Cornell University, Ithaca, NY, 14853}

\fntext[fn1]{Present address: Lawrence Berkeley National Laboratory, Berkeley, CA, 94720}

\begin{abstract}

We present the development and validation of a new multivariate
$b$ jet identification algorithm (``$b$ tagger'') used at the CDF
experiment at the Fermilab Tevatron. At collider experiments, $b$ taggers
allow one to distinguish particle jets containing $B$ hadrons
from other jets. Employing feed-forward neural network 
architectures, this tagger is unique in its emphasis on using
information from individual tracks.  This tagger not only contains the
usual advantages of a multivariate technique such as maximal use of
information in a jet and tunable purity/efficiency operating points,
but is also capable of evaluating jets with only a single track. To
demonstrate the effectiveness of the tagger, we employ a novel method
wherein we calculate the false tag rate and tag efficiency as a function
of the placement of a lower threshold on a jet's neural network output
value in $Z+1$ jet and $t\bar{t}$ candidate samples, rich in
light flavor and $b$ jets, respectively.

\end{abstract}


\end{frontmatter}

\section{Introduction}
\label{sec:intro}
The identification of jets originating from $b$ quarks is an important
part of many analyses at high-energy physics colliders. Searches for
the Higgs boson and measurements of top-quark properties depend on the
ability to identify $b$ jets properly. Furthermore, in many new
physics models, the third generation holds a special role, and
therefore final states with $b$-quark jets are common. The high
momentum of $B$ hadrons coupled with their long lifetimes result in a
large distance between the interaction point and the decay vertex of the $B$ hadron (decay
length). Additionally, a 
significant fraction 
($\approx$20\%) of $B$ hadrons decay with a soft lepton, i.e., a
charged lepton with a few GeV of momentum. These qualities are key to
distinguishing $b$-quark jets from other types of jets.


Almost all information as to whether or not a given jet originates from a
$B$-hadron decay is carried in the tracks its charged particles leave in
the detector. There are a few salient features of $B$-hadron decays
which can be searched for via the tracks in a jet. The lifetime
of a $B^{0}$ ($B^{\pm}$, $\Lambda_{b}$) hadron is 1.52 ps (1.64 ps, 1.42 ps). The distances these
particles travel during their lifetimes can be resolved by 
the 
CDF 
tracking system, and it is therefore possible to identify the delayed
decay of a $B$ hadron through the displacement of individual tracks
with respect to the primary interaction point (the primary vertex) and also through the
combining of tracks in the form of a fitted secondary decay vertex. 
Due to the large mass of the $b$ quark, the decay
products of $B$ hadrons will form a larger invariant mass than those
of hadrons not containing $b$ quarks. Furthermore, the large relativistic
boost typical of a $B$ hadron will result in decay products which tend to be more
energetic and collimated within a jet cone than other particles. Finally, particle
multiplicities tend to be different for jets containing $B$-hadron
decays compared to other jets; in particular, muons and electrons
appear in approximately 20\% of jets containing a $B$ hadron,
typically either directly via semileptonic decay of the $B$ or
indirectly through the semileptonic decay of a $D$ or $\Lambda_c$
resulting from a $B$ decay.


Many algorithms used at CDF were instrumental in the 1995 discovery of
the top quark~\cite{cdftop_evidence}.  Here we review the standard
$b$-tagging algorithms used at CDF.  Similar techniques as those
described in this paper have been developed at the D0
experiment~\cite{d0tagging} and at the CMS and ATLAS experiments at
the LHC~\cite{CMS-PAS-BTV-11-001,ATLAS-CONF-2011-102}.

SecVtx~\cite{secvtx} is a secondary vertex tagger. It is the most
commonly used $b$ tagger at CDF. Using only significantly displaced
tracks that pass certain quality requirements within each jet's cone,
an iterative method is used to fit a secondary vertex within
the jet.  Given the relatively long lifetime of the $B$ hadron, the
significance of the two-dimensional decay length $L_\text{xy}$ in the
$r$-$\phi$ plane is used to select $b$-jet candidates. The algorithm can be
performed with different sets of track requirements and threshold
values. In practice, three operating points are used, referred
to as ``loose'', ``tight'', and ``ultra tight''.

The jet probability~\cite{JetProb} tagger on the other hand does not
look for a secondary vertex, but instead uses the distribution of the
impact parameter significance of tracks in a jet, where impact
parameter significance is defined as the impact parameter divided by
its measured uncertainty ($d_{0}/\sigma_{d_{0}}$).  By comparing these
values to the expected distribution of values from light jets, it is
possible to determine the fraction of light jets whose tracks would
be more significantly displaced from the primary vertex than those of
the jet under study. While light-flavor jets should yield a fraction
uniformly distributed from 0 to 1, due to the long $B$ lifetime,
$b$ jets often produce significantly displaced tracks and hence tend
toward a fraction of 0. Although this algorithm produces a
continuous variable for discriminating $b$ jets, in practice only
three operating points are supported (jet probability $<$ 0.5\%, 1\%,
and 5\%).

Soft-lepton taggers~\cite{slt} take a different approach to $b$
tagging. Rather than focusing on tracks within a jet, they identify
semi-leptonic decays by looking for a lepton matched to a jet.  The
branching ratio of approximately 10\% per lepton makes this method
useful; however, if used alone, this class of tagger is not
competitive with the previously mentioned taggers. Because a
soft-lepton tagger does not rely on the presence of displaced tracks
or vertices, it has a chance to identify $b$ jets that the other
methods cannot.  In practice in CDF only the soft muon tagger is used
since high-purity electron or tau identification within jets is very
difficult.

Neural networks (NNs) can use as many flavor discriminating
observables as is computationally feasible; hence the efficiency of NN
taggers is often equal to or greater than that of conventional taggers
for a given purity. One such NN-based algorithm at CDF, called the ``KIT flavor 
separator"~\cite{Richter:2007zzc}, analyzes SecVtx-tagged jets and
identifies secondary vertices that are likely from long-lived 
$B$ hadrons, separating them from jets with secondary vertices that  
originate from charm hadrons or that are falsely reconstructed. This flavor 
separator has been used in many CDF analyses, notably in the CDF observation
of single top quark production~\cite{Aaltonen:2010jr}. Another NN-based algorithm, the ``Roma
tagger"~\cite{ferrazza06,Mastrandrea:2008zz}, has been used at CDF in
light Higgs searches. While the SecVtx 
tagger attempts to find exactly one displaced vertex in a jet, the
Roma tagger uses a vertexing algorithm that can find multiple
vertices, as may be the case when multiple hadrons decay within the
same jet cone (for example, in a $B \rightarrow D$ decay).
Three types of NNs are used: one to distinguish heavy from
light vertices, another to distinguish heavy-candidate from light-candidate
unvertexed tracks, and a third that takes as inputs the first two NN
outputs along with other flavor discriminating information, including
SecVtx and jet probability tag statuses, number of identified muons,
and vertex displacement and mass information. The performance of the
Roma tagger is roughly equivalent to SecVtx at its operating points
but allows for an ``ultra loose'' operating point yielding greater
efficiency, useful in certain analyses.

In this paper we describe a new tagger that builds on the development
of these taggers using feed-forward NN architectures.  The NNs provide
the ability to exploit correlations in many variables.  The tagger is
unique in its emphasis on individual tracks, and in its ability to
evaluate jets with only a single track. Each track's potential for
having come from a $B$-hadron decay is evaluated by a NN, and the
outputs of this NN are fed into a jet-wide NN along with other jet
observables such as the significance of the displacement of the
secondary vertex. The output of this NN, which we call the jet
$b$ness, is designed to identify jets containing a $B$-hadron decay.
The continuity of the NN output value allows for a tunable operating
point corresponding to the desired purity and efficiency.  

To characterize the tagger's performance, the efficiency and mistag
rate are obtained as a function of the jet $b$ness cut in $Z+1$ jet
(rich in light flavor jets) and $t\bar{t}$ (rich in $b$ jets)
candidate samples.  This choice of data samples differs from many
previous evaluations of performance using generic di-jet samples. The
large data sample accumulated at the Tevatron allow us to use the more
pure top quark samples for $b$ tagging efficiency studies.  The
ultimate use of this tagger is aimed at searches for standard model
dibosons and Higgs bosons. The momentum spectrum of $b$ quarks in top
pair production is better matched to these searches than the
relatively soft quark momentum spectrum found in generic di-jet
samples.  Finally, since our tagger will incorporate information from
many different tagging methods, techniques that, for instance, use soft
lepton-tagged jets as an input to an efficiency measurement for
displaced-vertex taggers cannot be used.


\section{The CDF Detector}
\label{sec:detector}

The CDF~II detector is described in detail
elsewhere~\cite{CDF_detect_A}.  The detector is cylindrically
symmetric around the proton beam line\footnote{%
  The proton beam direction is defined as the positive $z$
  direction.  The polar angle, $\theta$, is measured from the origin of
  the coordinate system at the center of the detector with respect to
  the $z$ axis, and $\phi$ is the azimuthal angle. Pseudorapidity,
  transverse energy, and transverse momentum are defined as
  $\eta$=$-\ln\tan(\theta/2)$, $E_{T}$=$E\sin\theta$, and
  $p_{T}$=$p\sin\theta$, respectively.  The rectangular coordinates $x$
  and $y$ point radially outward and vertically upward from the Tevatron
  ring, respectively.}
with tracking systems that sit within a superconducting solenoid which
produces a $1.4$~T magnetic field aligned coaxially with the $\ppbar$
beams. The Central Outer Tracker (COT) is a $3.1$ m long open cell
drift chamber which performs 96 track measurements in the region
between $0.40$ and $1.37$ m from the beam axis, providing coverage in
the pseudorapdity region $|\eta| \le 1.0$~\cite{cot_nim}. Sense wires
are arranged in eight alternating axial and $\pm2^{\circ}$ stereo
``superlayers" with 12 wires each. The position resolution of a single
drift time measurement is about $140~\mu$m.

Charged-particle trajectories are found first as a series of
approximate line segments in the individual axial superlayers. Two
complementary algorithms associate segments lying on a common circle,
and the results are merged to form a final set of axial tracks. Track
segments in stereo superlayers are associated with the axial track
segments to reconstruct tracks in three dimensions. 


The efficiency for finding isolated high-momentum tracks is measured
using electrons from $W ^{\pm} \rightarrow e^{\pm} \nu$ decays
identified in the central region $|\eta| \le 1.1$ using only
calorimetric information from the electron shower and the missing
transverse energy. In these events, the efficiency for finding the
electron track is $99.93^{+0.07}_{-0.35}\%$, and this is typical for
isolated high-momentum tracks from either electronic or muonic $W$
decays contained in the COT. The transverse momentum resolution of
high-momentum tracks is $\delta p_T/ p_T^2 \approx 0.1\% 
\,({\rm GeV}/c)^{-1}$. Their track position resolution in the direction
along the beam line at the origin is $\delta z \approx 0.5$\,cm, and
the resolution on the track impact parameter, the distance from the
beam line to the track's closest approach in the transverse plane, is
$\delta d_{0} \approx 350~\mu$m.


A five layer double-sided silicon microstrip detector (SVX) covers
the region between $2.5$ to $11$ cm from the beam axis. Three separate
SVX barrel modules along the beam line cover a
length of 96 cm, approximately 90\% of the luminous beam interaction
region. Three of the five layers combine an $r$-$\phi$ measurement on
one side and a $90^{\circ}$ stereo measurement on the other, and the
remaining two layers combine an $r$-$\phi$ measurement with small angle stereo at
$\pm1.2^{\circ}$. The typical silicon hit resolution is 11~$\mu$m. Additional Intermediate Silicon Layers (ISL) at radii between 19
and 30 cm from the beam line in the central region link tracks in the
COT to hits in the SVX.

Silicon hit information is added to COT tracks using a
progressive ``outside-in" tracking algorithm in which COT tracks are
extrapolated into the silicon detector, associated silicon hits are
found, and the track is refit with the added information of the
silicon measurements. The initial track parameters provide a width for
a search road in a given layer. Then, for each candidate hit in that
layer, the track is refit and used to define the search road into the
next layer. This stepwise addition of precision SVX information at
each layer progressively reduces the size of the search road, while
also accounting for the additional uncertainty due to multiple
scattering in each layer. The search uses the two best candidate hits
in each layer to generate a small tree of final track candidates, from
which the tracks with the best $\chi^{2}$ are selected. The efficiency
for associating at least three silicon hits with an isolated COT track
is $91 \pm 1\%$. The extrapolated impact parameter resolution for
high-momentum outside-in tracks is much smaller than for COT-only
tracks: $30~\mu$m, including the uncertainty in the beam position.

Outside the tracking systems and the solenoid, segmented calorimeters
with projective geometry are used to reconstruct electromagnetic (EM)
showers and jets. The EM and hadronic calorimeters are
lead-scintillator and iron-scintillator sampling devices,
respectively. The central and plug calorimeters are segmented into
towers, each covering a small range of pseudorapidity and azimuth,
and in full cover the entire $2\pi$ in azimuth and the pseudorapidity
regions of $|\mathrm\eta|$$<$1.1 and 1.1$<$$|\mathrm\eta|$$<$3.6
respectively. The transverse energy $E_T$, where the
polar angle is calculated using the measured $z$ position of the event
vertex, is measured in each calorimeter tower. Proportional and
scintillating strip detectors measure the transverse profile of EM
showers at a depth corresponding to the shower maximum.

High-momentum jets, photons, and electrons leave isolated energy
deposits in contiguous groups of calorimeter towers which can be
summed together into an energy cluster. 
Electrons are identified in the central EM calorimeter as isolated,
mostly electromagnetic clusters that also match with a track in the
pseudorapidity range $|\eta| < 1.1$. The electron transverse energy is
reconstructed from the electromagnetic cluster with precision
$\sigma(E_{T})/E_{T} = 13.5\% / \sqrt{E_{T} ({\rm GeV})} \oplus 2\%$,
where the $\oplus$ symbol denotes addition in quadrature. Jets are
identified as a group of electromagnetic and hadronic calorimeter
clusters using the \textsc{jetclu} algorithm~\cite{jetclu} with a cone
size of 0.4. Jet energies are corrected for the calorimeter
non-linearity, losses in the gaps betwen towers, multiple primary
interactions, the underlying event, and out-of-cone
losses~\cite{jesnim}. The jet energy resolution is approximately
$\sigma_{E_{T}} = 1.0~{\rm GeV} + 0.1\times E_T$

Directly outside of the calorimeter, four-layer stacks of planar drift
chambers detect muons with $p_{T} > 1.4~$GeV$/c$ that traverse 
the five absorption lengths of the calorimeter. Farther out, behind an
additional 60 cm of steel, four layers of drift chambers detect muons
with $p_{T} > 2.0~$GeV$/c$. The two systems both cover a region of
$|\eta| \le 0.6$, though they have different structure and their
geometrical coverages do not overlap exactly. Muons in the region
between $0.6 \le |\eta| \le 1.0$ pass through at least four drift
layers lying in a conic section outside of the central
calorimeter. Muons are identified as isolated tracks in the COT that
extrapolate to track segments in one of the four-layer stacks.

\section{Description of the neural network}

All neural networks are trained using simulated data samples. The
geometric and kinematic acceptances are obtained using a {\sc
geant}-based simulation of the CDF II detector~\cite{geant3}. For the
comparison to data, all sample cross sections are normalized to the
results of NLO calculations performed with the {\sc mcfm~v5.4}
program~\cite{campbell} and using the \textsc{cteq6m} parton
distribution functions~\cite{cteq6m}.


\subsection{Basic track selection}

A great deal of information as to whether a jet contains a $B$-hadron
decay is contained within the jet's individual tracks. Indeed, as
described earlier, the jet probability algorithm~\cite{JetProb} uses
information solely based on the significance of the impact parameters
of tracks. Furthermore, an important choice to make when seeking
displaced vertices is which tracks to use as candidates for a fit. In
light of this, our tagger takes a ground-up approach where
the first step in the evaluation of how $b$-like a jet is involves
using a neural network to discriminate $B$-hadron decay tracks
from other tracks in a jet. 
We use relatively loose criteria when selecting which
tracks to evaluate with our track-by-track NN,
thereby improving the $b$-tagging efficiency.
We reject tracks that use hits only in the COT, 
as the COT alone has insufficient resolution to
distinguish the effects of the displacement of a $B$-hadron decay from
the primary vertex. Additionally, a track must have a $p_T > 0.4~\mathrm{
GeV}/c$, a requirement CDF maintains for all tracks, and be found
within a cone of $\Delta R < $ 0.4 about the jet axis, where $\Delta R
= \sqrt{(\Delta \phi)^{2} + (\Delta \eta)^{2}}$. Finally, for tracks
within a jet, track pairs are removed if they are oppositely charged,
form an invariant mass within 10 MeV of that of a $K_S$ (0.497
GeV$/c^{2}$) or $\Lambda$ (1.115 GeV$/c^{2}$), and can be fit into a
two-track vertex.  This requirement is included to reject non-$b$ jets
that contain these long-lived particles, as they can mimic $b$ jets,
compromising our purity.

\subsection{The track neural network}
\begin{figure*}[tbhp]
  \centering
  \includegraphics[width=0.8\textwidth]{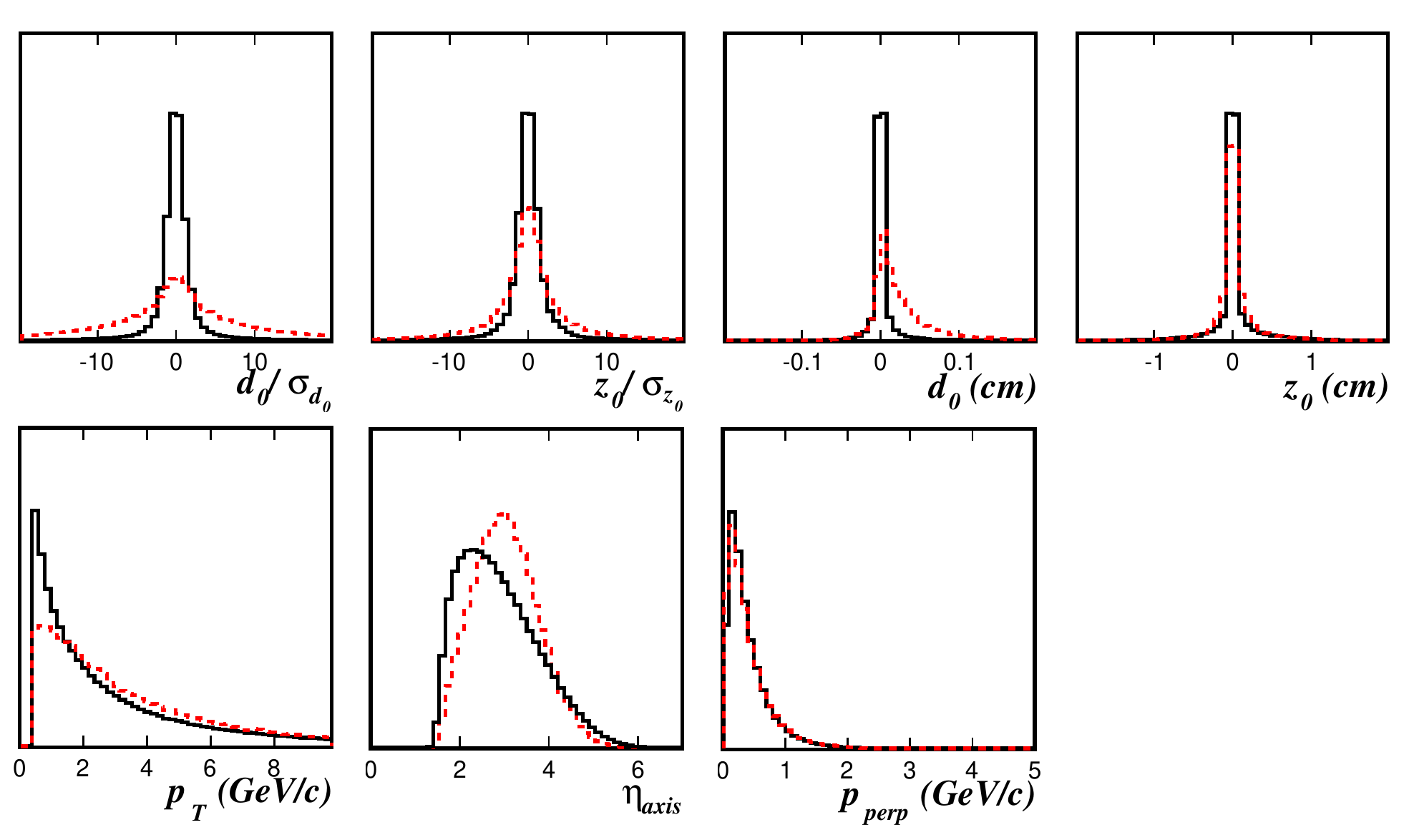}
  \caption{ Inputs used in the neural network for calculating the
    per-track $b$ness. The red dashed line is signal and the black solid line is
    background. The $y$-axis is in arbitrary units. } 
\label{fig:trkcomp_plot}
\end{figure*}

The two primary categories of input variables to the track-by-track NN
are observables related to the displacement of the track from the
primary vertex and observables related to the kinematics of the
track. The former category includes the track's signed impact
parameter\footnote{We define the signed impact parameter of a track as
  positive if the angle between the candidate b-jet direction and the
  line joining the primary vertex to the point of closest approach of
  the track to the vertex is less than $90^\circ$, and as negative
  otherwise.}  ($d_0$), its $z$ displacement ($z_0$) from the primary
vertex, and the significances of these two quantities, given their
uncertainties ($d_0/\sigma_{d_0}$ and $z_0/\sigma_{z_0}$). The latter
category takes advantage of the fact that tracks from $B$-hadron
decays have a somewhat harder $p_T$ spectrum than other tracks, and
are more collimated within a jet. This category includes the track's
$p_T$, its pseudorapidity ($\eta_\text{axis}$) with respect to the jet axis, 
and its momentum ($p_\text{perp}$) perpendicular to the jet axis.

A final input variable to the track-by-track $b$ness NN is the $E_T$ of the jet, 
since distributions of the track observables are correlated with
their parent jet $E_T$. 
To ensure that the distributions of track observables used to
train the track-by-track NN are not kinematically biased, $B$ hadron
and non-$B$ hadron tracks are weighted in training to have the
same parent jet $E_T$ distribution. 


Figure~\ref{fig:trkcomp_plot} shows distributions of the track
variables in \textsc{pythia}~\cite{pythia} $ZZ \rightarrow jjjj$ Monte
Carlo simulations (MC) for tracks matched by $\Delta R < 0.141$ to particles that
come from $B$-hadron decays compared to tracks in jets which are not
matched to $B$ hadrons. These figures indicate that the displacement
variables tend to give more discrimination power than the kinematic
variables; in particular, the impact parameter variables are the most
important inputs to the NN.
%
%

The NN is a feed-forward multilayer perceptron with a single output
and two hidden layers of 15 and 14 nodes implemented using the MLP
algorithm from the TMVA package~\cite{tmva}. The same number of signal
and background events was used in the training. The performance of the
NN was similar with larger numbers of hidden layer nodes.



\subsection{The jet neural network}


To determine how $b$-like a jet is, we train a NN to
distinguish jets containg $B$-hadron decays from those not containing
$B$-hadron decays. Many of the input variables come directly from the
track-by-track NN described in the previous section: the NN values of
the five most $b$-like tracks ($b_i$, $i = 0..4$), as well as the number
($n_\text{trk}$) of tracks with a NN output greater than 0.

We use tracks with track-by-track NN values greater than -0.5 in the fitting 
of a secondary vertex. An initial fit is performed with all such tracks;
if the largest contribution to the total fit $\chi^{2}$ from any of
them exceeds a value of 50, it is removed, and the remaining tracks
are re-fit. This process continues until either the largest $\chi^{2}$
contribution from any track is less than 50, or there are fewer than
two tracks to be fit. 
If a secondary vertex is successfully fit, then the significance of
its displacement from the primary vertex ($L_{xy}/\sigma_{L_{xy}}$) and
the invariant mass  ($m_\text{vtx}$) of the tracks used to fit it both
serve as inputs into the NN.

\begin{figure*}[tbp]
  \centering
  \includegraphics[width=0.8\textwidth]{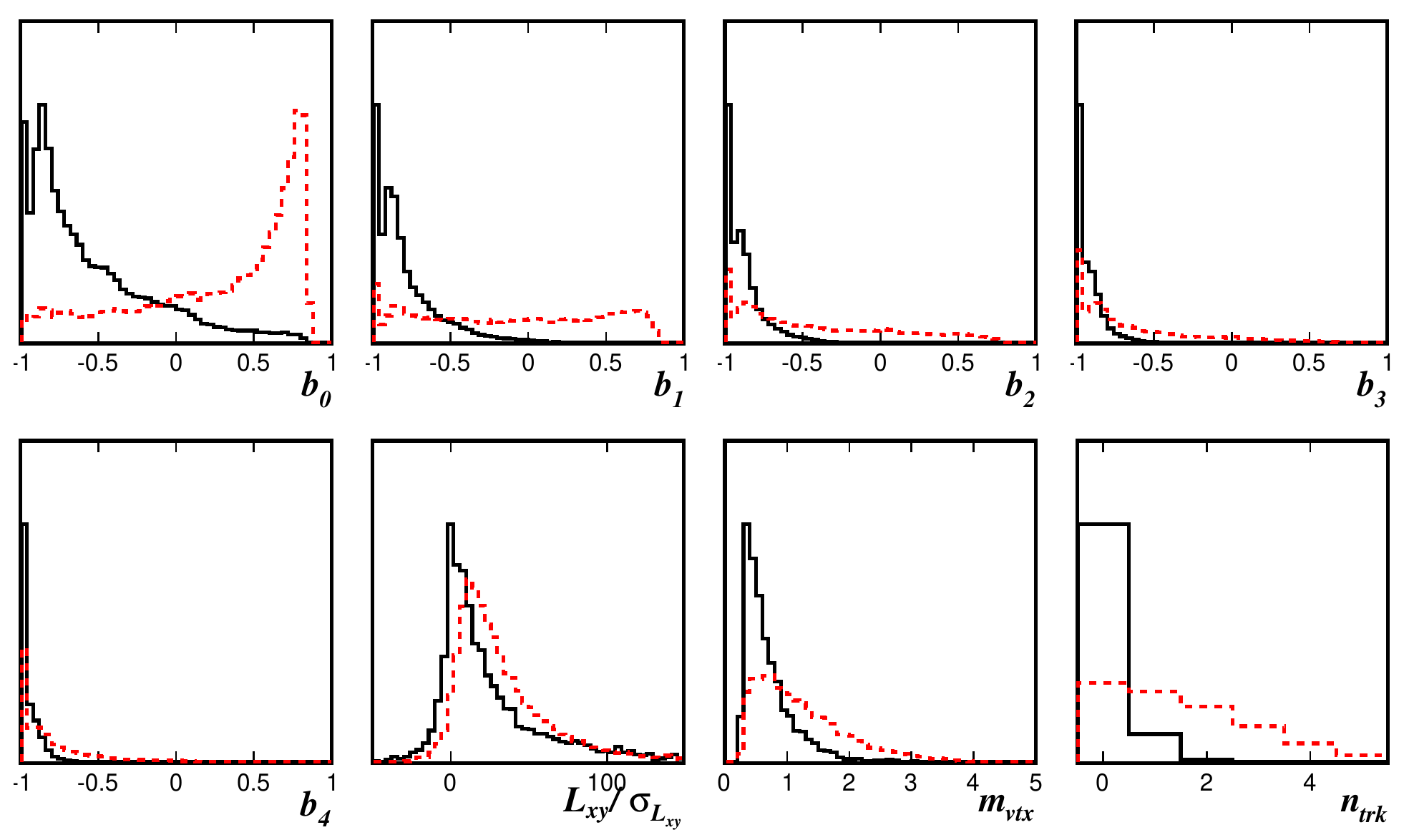}
  \caption{ The most relevant inputs used in the neural network for
    calculating the per-jet $b$ness. The red dashed line is signal and the
    black solid line is background. $b_i$ refers to the $b$ness of the
    $i$\textsuperscript{th} track, ordered in $b$ness. The $y$-axis is in arbitrary units.}
\label{fig:jetcomp_plot}
\end{figure*}
\begin{figure*}[tbp]
  \centering
  \includegraphics[width=0.8\textwidth]{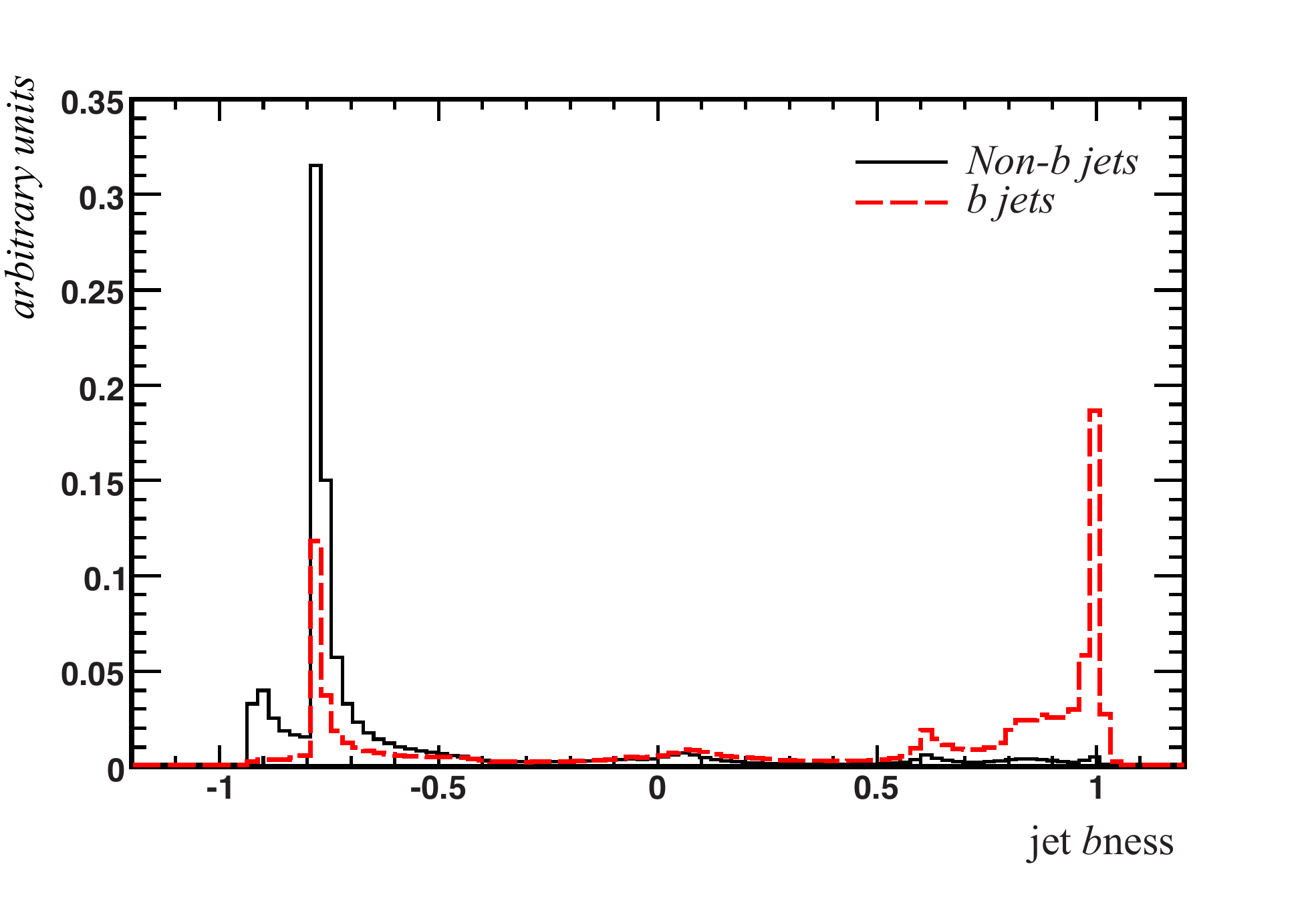}
  \caption{ Output of the final neural network, for signal (red dashed
    line) and background (black solid line). Good separation is seen
    with the exception of signal (and background) peaking near a jet
    $b$ness of -0.8.  This region is dominated by jets with zero
    tracks having positive track $b$ness, zero $K_S$ candidates found,
    and no secondary vertex.  Indeed, some $b$ jets are
    indistinguishable from non-$b$ jets.  The sharp features of this
    distribution are a result of the discrete inputs to the NN.}
\label{fig:bness_comp_both}
\end{figure*}
Additionally, because a much higher fraction of $b$ jets than non-$b$
jets contain $K_S$ particles, the number of $K_S$ candidates found is used as an
input to the jet-by-jet NN.  Finally, if there is a muon candidate in
the jet cone, its likelihood to be a true muon is used as an input.
This value is calculated using the soft muon tagger~\cite{slt}
described above. The architecture of the jet-by-jet NN is similar to
that of the track-by-track NN, with two hidden layers of 15 and 16
nodes.
As in the track NN, to avoid a kinematic bias, the parent jet $E_T$ distributions
are weighted to be equal and also input into the NN.
Distributions of the most important jet-by-jet NN input variables are
shown in Figure~\ref{fig:jetcomp_plot}. Distributions of the NN output
are shown in Figure~\ref{fig:bness_comp_both}.




The training for the track NN as well as the jet NN
is performed using jets, from a \textsc{pythia} $ZZ$ MC sample,
matched to $b$ quarks from
$Z\rightarrow b\bar{b}$ events for signal and jets not matched to $b$
quarks for background.
\section{Selection for Mistag Rate and Efficiency Determination}
\label{sec:selection}
\begin{table*}[tbph]
 \begin{minipage}{1.0\linewidth}
  \begin{center}
    \begin{tabular}{|c|} \hline
      \textit{$Z + 1$ jet Selection} \\ \hline\hline
      $N_\text{leptons} = 2$, both electrons or both muons \\
      Leptons have opposite charge \\
      $\Delta z_{0}$ between leptons $<$ 5 cm \\ 
      Lepton $p_{T} >$ 20 GeV/$c$ \\
      75 GeV/$c^{2} < M_{ll} <$ 105 GeV/$c^{2}$ \\
      $\mett < 25$ GeV \\
      Reconstructed $p_{T}(Z)>10$ GeV$/c$ \\
      $N_\text{jets}$($E_{T} > 10$ GeV) $= 1$ \\
      Jet $E_{T} > 20$ GeV, $|\eta| < 2.0$ \\\hline
    \end{tabular}
    \quad
    \begin{tabular}{|c|} \hline
      \textit{$t\bar{t}$ Selection}  \\ \hline\hline

      $N_\text{leptons} = 1$ \\
      Lepton $p_{T} >$ 20 GeV/$c$ \\
      $\mett > 20$ GeV \\
      $\mett$-significance $> 1(3)$ for $\mu(e)$ events \\
      Reconstructed $M_{T}(W) >$ 28 GeV/$c^2$ \\
      Highest two $b$ness jets' $E_{T} >$ 20 GeV \\
      $N_\text{jets}(E_{T} > 15$ GeV) $\ge 4$ \\
      Total sum $E_{T} > 300$ GeV \\
\\
      \hline
    \end{tabular}
   \end{center}
 \end{minipage}
\caption{Summary of event selection requirements for the $Z + 1$ jet and
  $t\bar{t}$ samples. The total sum $E_{T}$ is defined as the sum
  of the lepton $p_{T}$, $\mett$, and $E_{T}$ of all jets with $E_{T}
  > $ 15 GeV.}
\label{tab:selection}

\end{table*}

In order to use this new $b$ tagger in analyses, 
we determine the efficiency and false tag
(``mistag'') rate as a function of a minimal $b$ness requirement, $e(b)$ and
$m(b)$ respectively. We use comparisons between data and Monte Carlo
simulation to evaluate these quantities and their uncertainties. Also, we evaluate 
the efficiency and mistag rate in Monte Carlo  ($e_\text{MC}(b)$ and $m_\text{MC}(b)$, 
respectively), and determine the necessary scale factor, $s_{e}(b) = e(b)/e_\text{MC}(b)$ (with 
a similar definition for the mistag rate), to correct the simulation.

\begin{table*}[tbh]
\begin{center}
\begin{tabular}{rr@{$\,\pm\,$}rr@{$\,\pm\,$}r} \hline
& \multicolumn{2}{c}{Electrons} & \multicolumn{2}{c}{Muons} \\ \hline \hline
\textit{$Z + 1$ jet selection}  \\ \hline
Data Events & \multicolumn{2}{l}{9512} & \multicolumn{2}{l}{5575}\\ 
MC Events & 9640 & 880& 5540 & 490\\ \hline \hline
\textit{$t\bar{t}$ Selection} \\ \hline
Data Events & \multicolumn{2}{l}{~~507} & \multicolumn{2}{l}{~~835} \\ 
MC Events & 542 & 56 & 862 & 85\\ \hline
\end{tabular}
\caption{Number of events in data and MC in the $Z + 1$ jet selection
  region, after proper scale factors have been applied. The
  uncertainties on the MC reflect only the two dominant systematic
  uncertainties: the uncertainty on the jet energy scale and the
  uncertainty on the luminosity. Overall, the agreement in number of
  events is good.}
\label{tab:events}
\end{center}
\end{table*}
Following the procedure described in~\ref{sec:equations} and~\ref{sec:equations2}, we must choose two
independent regions in which to determine the mistag rate and efficiency
of the $b$ tagger. To reduce uncertainties, it is best to choose a
well-modelled region dominated by falsely tagged jets (where we expect few
$b$ jets) and a well-modelled region rich in $b$ jets. For the former,
we choose events containing two oppositely charged electrons or muons
likely from the decay of a $Z$ boson, plus one jet. For the latter, we
choose events containing the decay of a pair of top quarks, where we
require exactly one lepton, at least four jets, and a large imbalance
in transverse momentum in the event, indicating the likely presence of
a neutrino. We expect that the two jets with the highest $b$ness values in
this sample will very likely be $b$ jets.  The cuts applied for these
two selection regions are described in Table \ref{tab:selection}. We
use the $\mett$ significance, as defined in \cite{metjj_prl,
emtiming}, to reduce any contribution from multi-jet production where
a jet is mis-identified as an electron or muon.\footnote{
  We define the missing transverse momentum $\vec{\mett}$$\equiv
  -\sum_i E_{\rm T}^i {\bf n}_i$, where ${\bf n}_i$ is the unit vector
  in the azimuthal plane that points from the beamline to the $i$th
  calorimeter tower. We call the magnitude of this vector \mett.  The
  \mett significance is a measure of the ratio of the value of \mett
  to its uncertainty, and tends to be small for \mett due to
  mismeasurement rather than due to undetected, long-lived neutral
  particles such as neutrinos.   
}

These events are selected by high-$p_{T}$ electron and
muon triggers. We use data corresponding to an integrated luminosity
of 4.8 fb\textsuperscript{-1}. We use \textsc{alpgen}~\cite{alpgen}, 
interfaced with \textsc{pythia} for parton showering, to model $W$ and $Z$
plus jets samples and \textsc{pythia} to model $t\bar{t}$ and other
processes with small contributions. We check the trigger efficiency against a sample
of $Z\to e^+e^-$ or $\mu^+\mu^-$ events without jets. Table~\ref{tab:events}
contains a summary of the total number of events.

\section{Mistag Rate Determination}
\label{sec:mistag}

\begin{figure*}[tbp]
  \centering
  \includegraphics[width=0.7\textwidth]{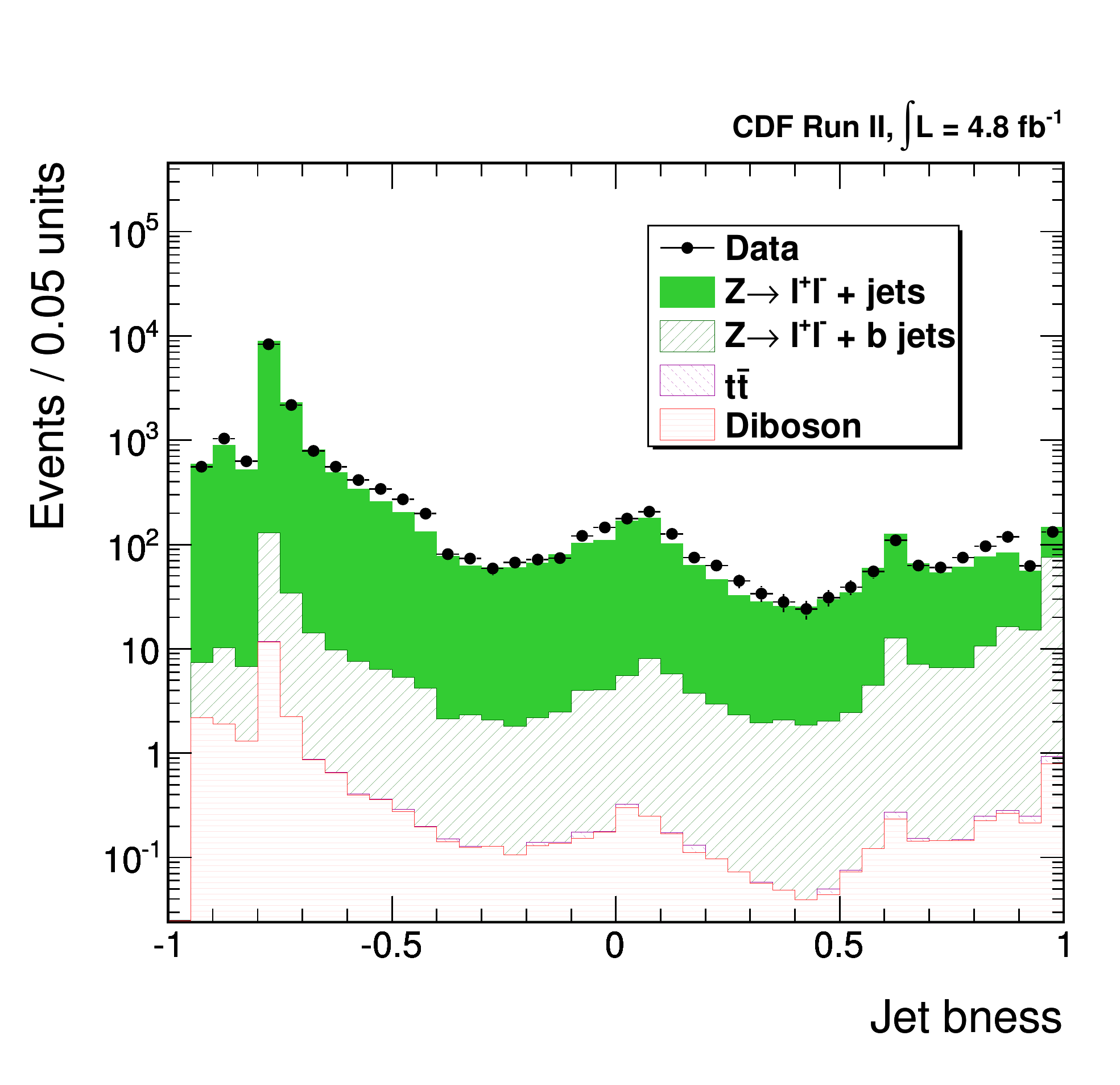}
\caption{A comparison of the jet $b$ness in data and MC in the $Z + 1$ jet 
selection region. The MC is able to reproduce the main features of the $b$ness distribution in data. We use this distribution to determine the mistag rate for placing a cut on jet $b$ness in data, and use the differences between data and MC to determine corrections to the mistag rate in MC.
}
\label{fig:z1jet-jet_bness}
\end{figure*}

\begin{figure*}[tbp]
  \centering
  \includegraphics[width=0.45\textwidth]{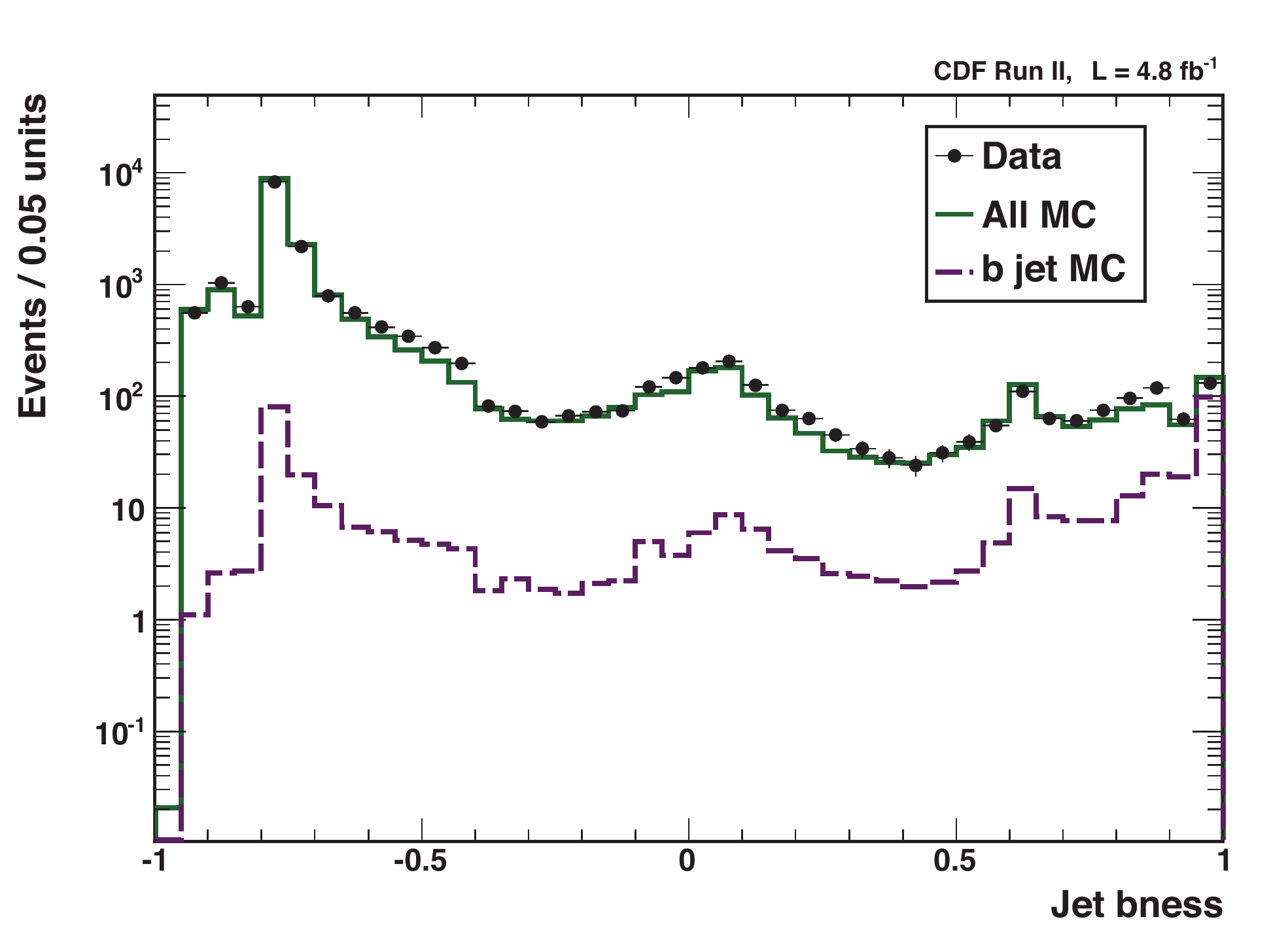}
  \includegraphics[width=0.45\textwidth]{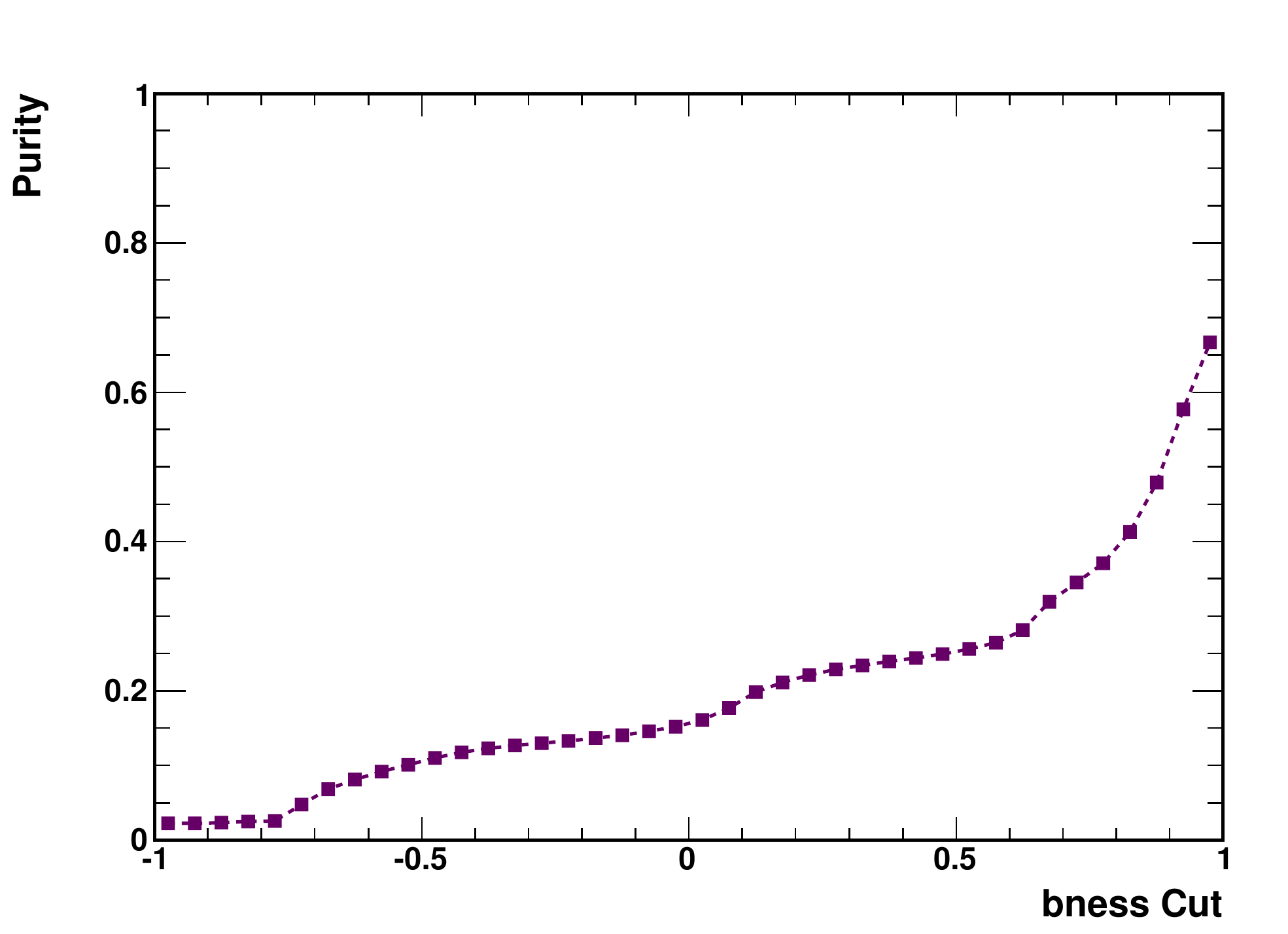}
\caption{\textit{Left}: A comparison of the jet $b$ness in data (black points) 
and MC (green solid line) in the $Z + 1$ jet selection region, with the portion of the MC jets 
matched to $b$ quarks (purple dashed line) shown independently. \textit{Right}: The $b$-jet 
purity for a given $b$ness cut, as determined from matched jets in the MC.
As we wish to use the $Z + 1$ jet sample as a model for mistags, it is necessary to 
subtract the significant $b$-jet contribution at high $b$ness values.}
\label{fig:z1jet-jet_bness-Matches}
\end{figure*}

Figure~\ref{fig:z1jet-jet_bness} shows the jet $b$ness distribution
for jets in the $Z$ + 1 jet sample. The sample is dominated
by light-flavor jets, but there is a significant contribution of real
$b$ jets at higher $b$ness values, coming from $Z+b\bar{b}$
production. This is seen more clearly in
Figure~\ref{fig:z1jet-jet_bness-Matches}, where we separate the MC
jets based on whether there are generator-level $b$ quarks located
within each jet's cone ($\Delta R = 0.4$). Also shown is the $b$-jet
purity  ($N_\text{b-jets} / N_\text{jets}$) as a function of lower threshold on jet $b$ness. We see the
$b$-jet incidence rate reaches above 60\% for the highest $b$ness
cuts, and thus we will expect the uncertainties in the mistag rate to
be substantially higher there, due to both the small
sample of available jets and the high contamination rate
combined with the uncertainty on the number of $b$ jets
in that smaller sample.

\begin{figure*}[tbp]
  \centering
  \includegraphics[width=0.48\textwidth]{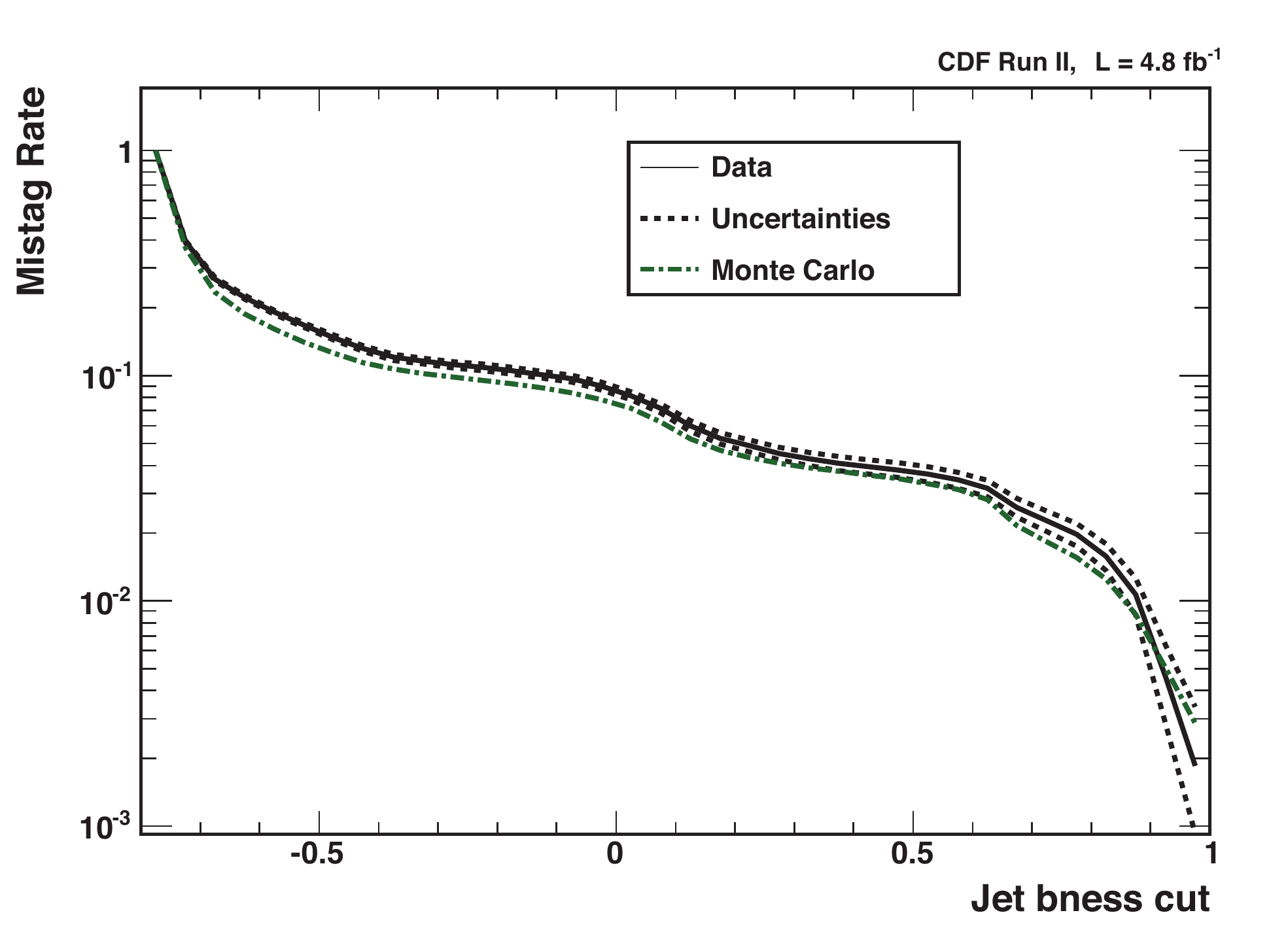}
  \includegraphics[width=0.48\textwidth]{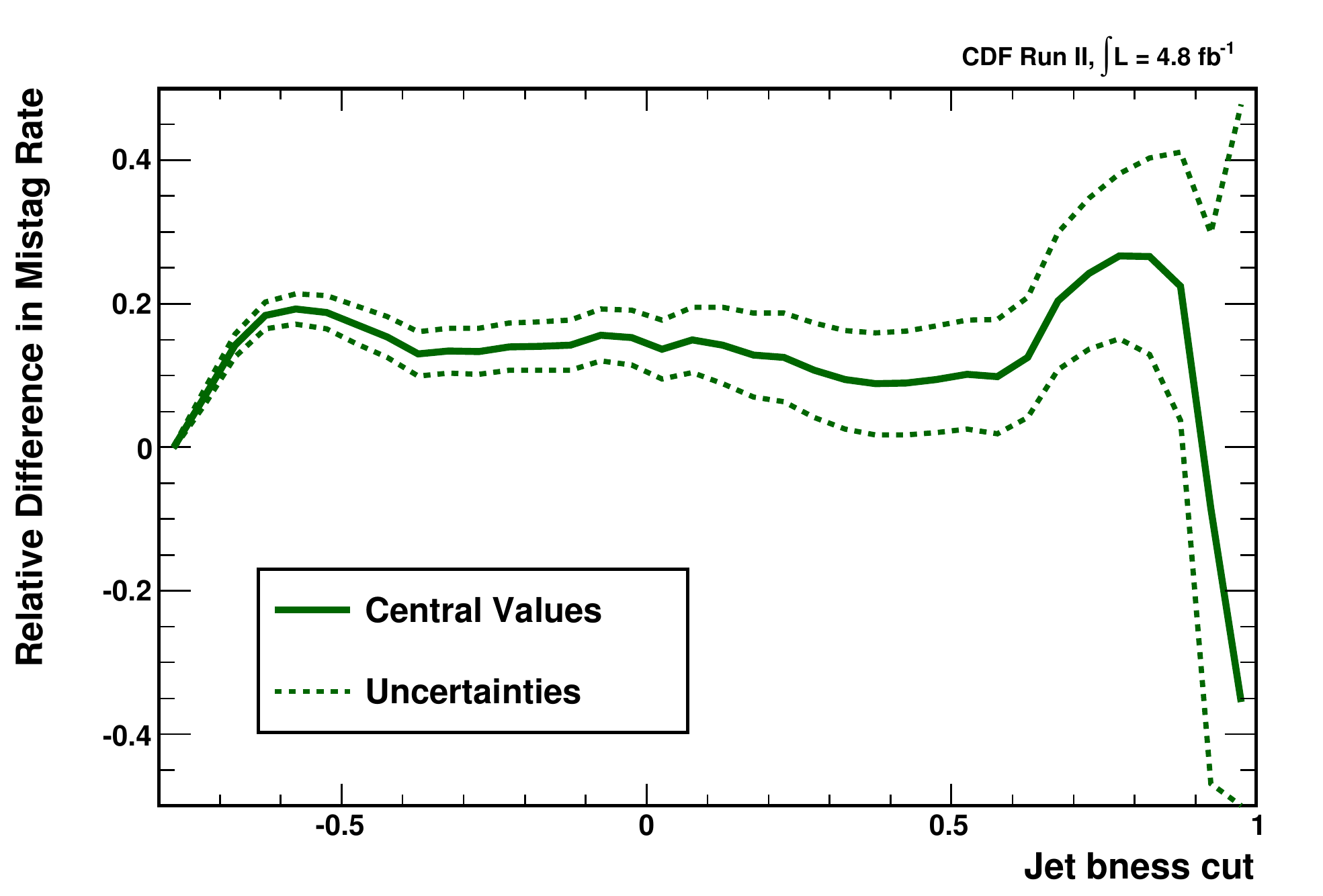}
\caption{\textit{Left}: The mistag rate in data (solid black line, dashed lines 
represent uncertainty) and Monte Carlo (dot-dashed green line) as a function the jet 
$b$ness.  We see our simulation typically under-predicts the mistag rate 
measured in data, requiring us to consider a correction to apply to the MC. 
\textit{Right}: The calculated MC scale factor on the mistag rate (solid line) 
and its uncertainty (dashed lines) relative to the mistag rate in the Monte 
Carlo. The value of the scale factors and their uncertainties at the relevant $b
$ness cuts in this analysis are summarized in Table \ref{tab:btag-values}. We 
see very large uncertainties on the mistag rate scale factor around the high jet 
$b$ness cut of 0.85, due to the small number of events and significant heavy-flavor
removal that must be done in this region.}
\label{fig:jet_bness-mistag}
\end{figure*}

The mistag rate for jets above a given $b$ness threshold is simply the fraction of non-$b$ jets 
above that threshold. To obtain this quantity, we use the fraction of jets in data above that threshold 
($m_\text{raw}(b)$), but must correct this quantity for the expected number of $b$ jets in our $Z + 1$ 
jet sample. We obtain an estimate of this $b$ jet contamination from MC simulation, and obtain the 
corrected mistag rate, $m(b)$.
We show the values 
of $m(b)$ as well as the relative difference between the 
mistag rate in data and MC ($s_{m}(b) - 1$) in Figure~\ref{fig:jet_bness-mistag}.

We can also calculate the uncertainty on the mistag rate given the error on the $b$-tagging efficiency 
and the uncertainty on the fraction of $b$ jets in our $Z + 1$ jet sample. The 
former is determined through iterative calculations incorporating the $t\bar{t}$ 
selection, while the latter we take to be 20\%~\cite{ZbPRD}. The resulting uncertainties are 
also shown in Figure~\ref{fig:jet_bness-mistag}.

\section{Tagging Efficiency Determination}
\label{sec:efficiency}

\begin{figure*}[tbp]
  \centering
  \includegraphics[width=0.45\textwidth]{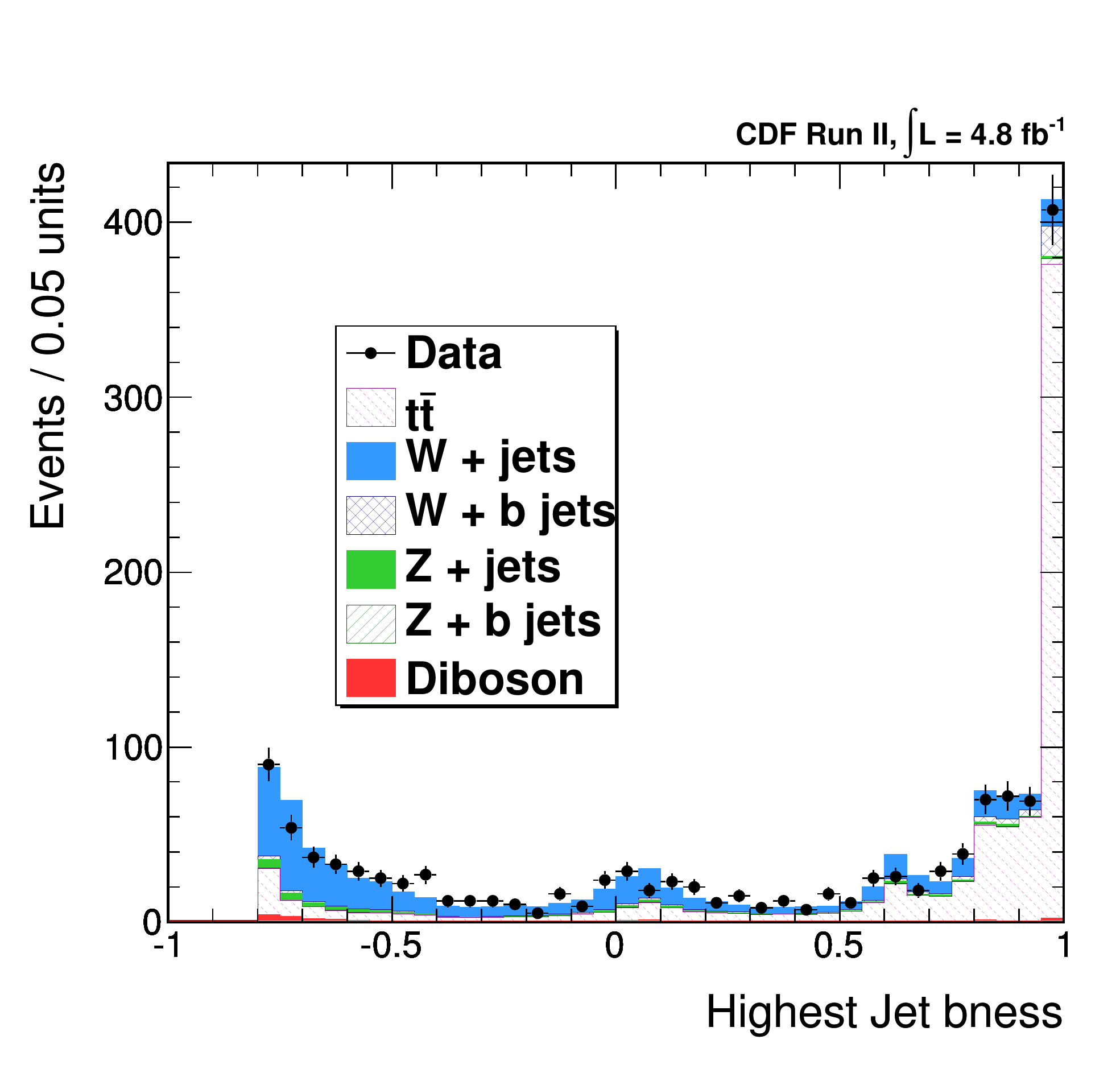}
  \includegraphics[width=0.45\textwidth]{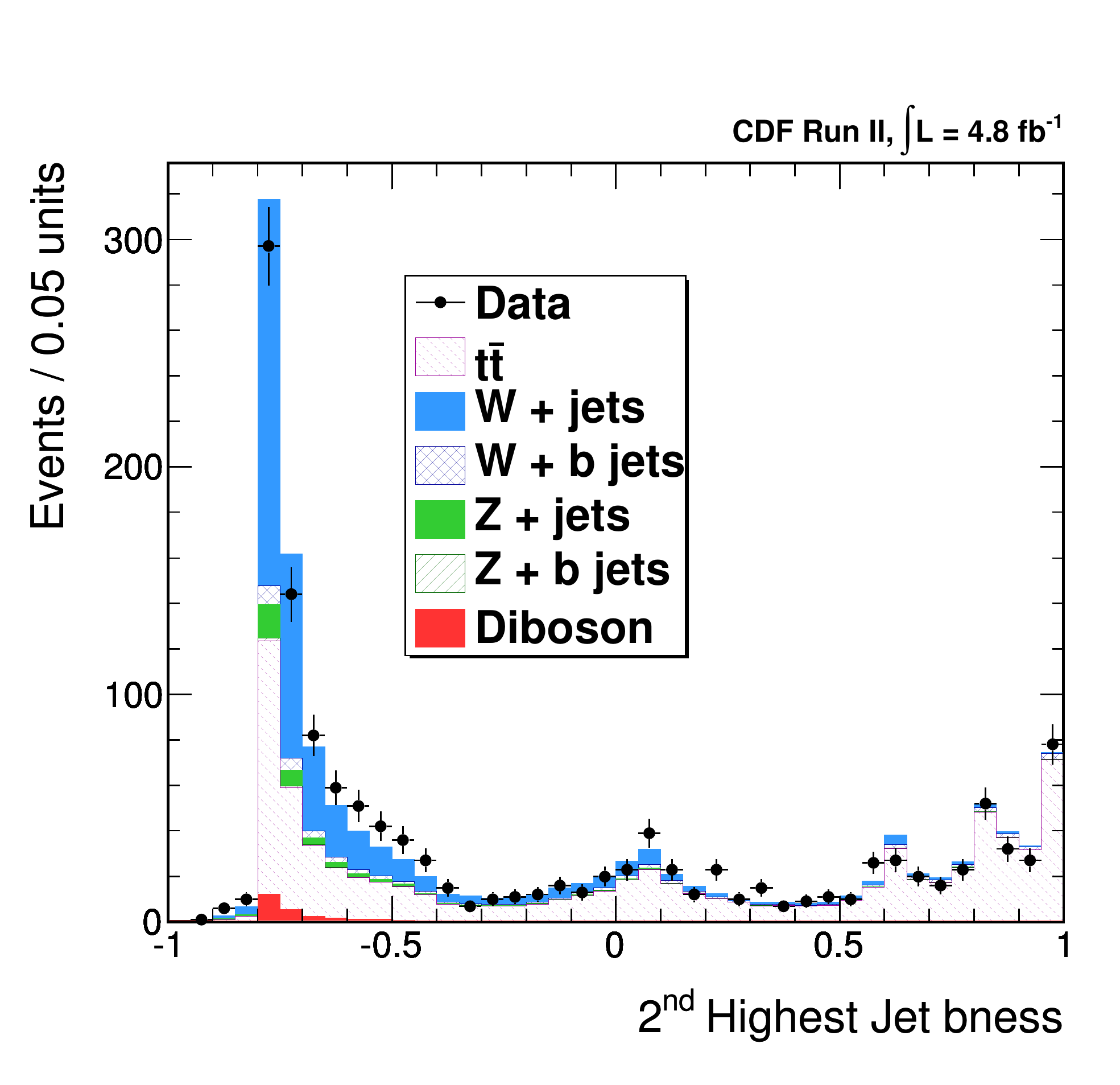}
  \caption{Jet $b$ness of the first (left) and second (right) jet, as
    ordered by $b$ness, in the $t\bar{t}$ lepton + jets selection
    region. The simulation reproduces most of the features of the
    data, and we see much of the $b$-enriched samples clustered
    towards high $b$ness.}
\label{fig:w2jet-jet_bness-Cut2}
\end{figure*}

\begin{figure*}[tbp]
  \centering
  \includegraphics[width=0.48\textwidth]{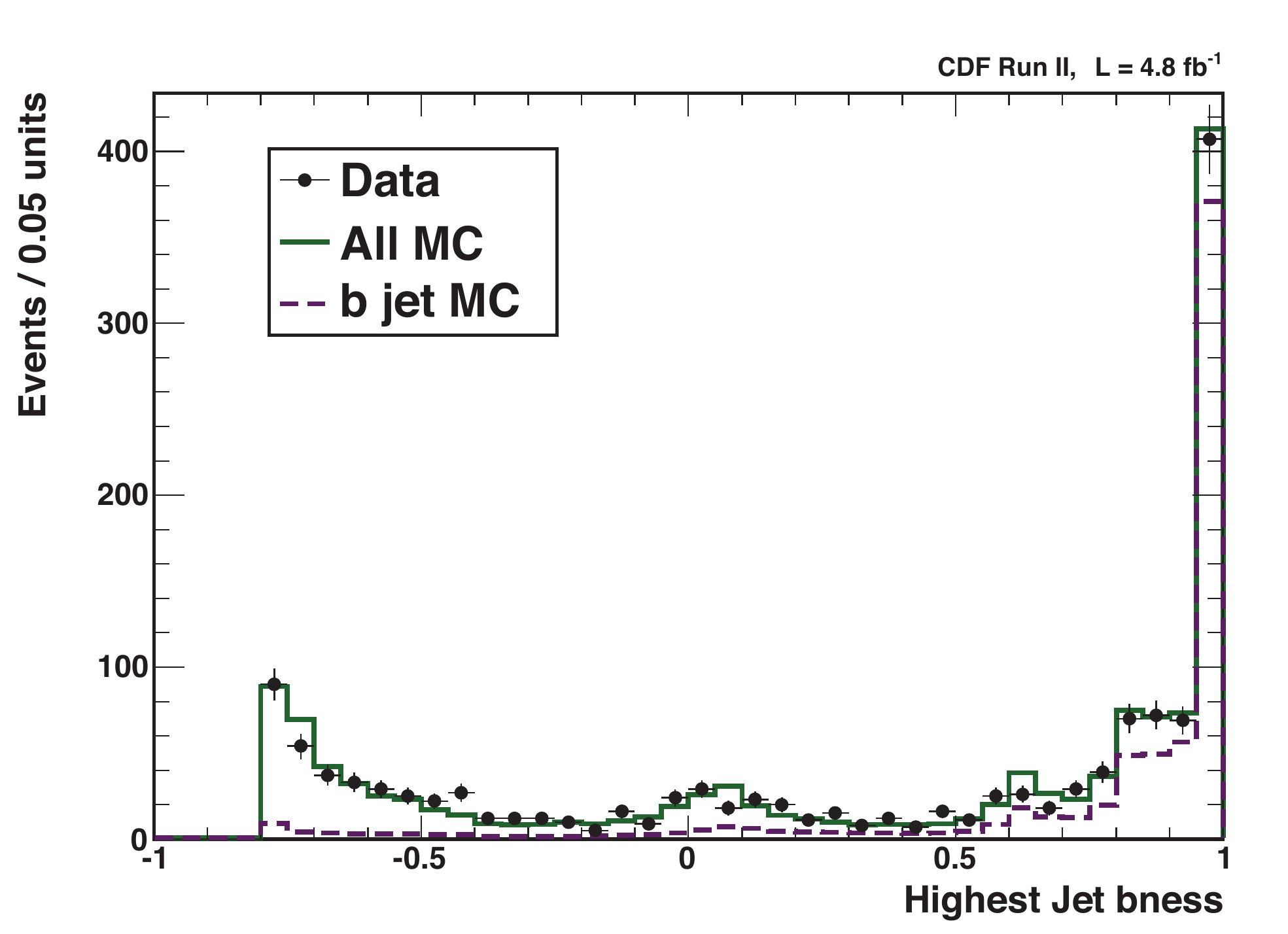}
  \includegraphics[width=0.48\textwidth]{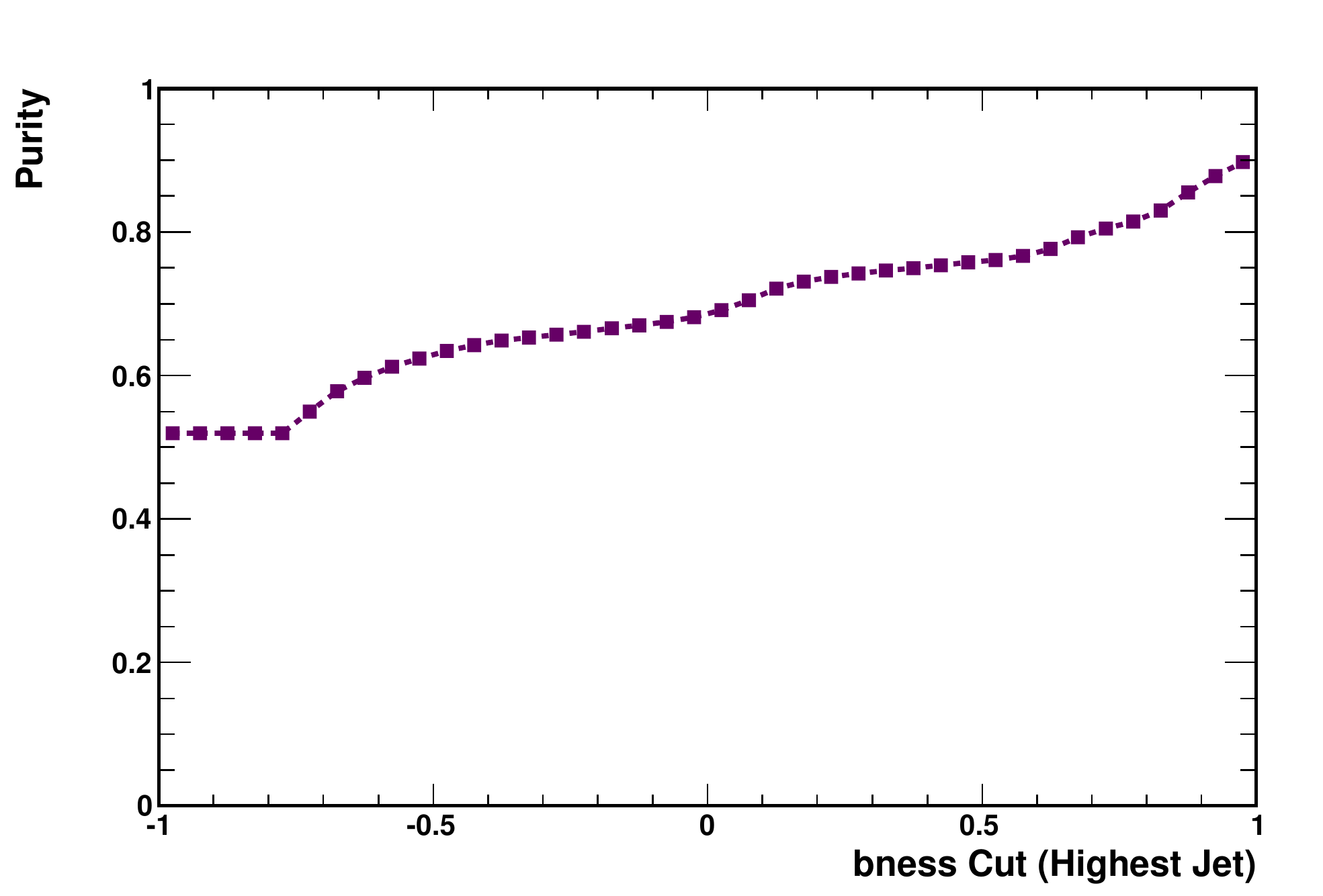}
  \includegraphics[width=0.48\textwidth]{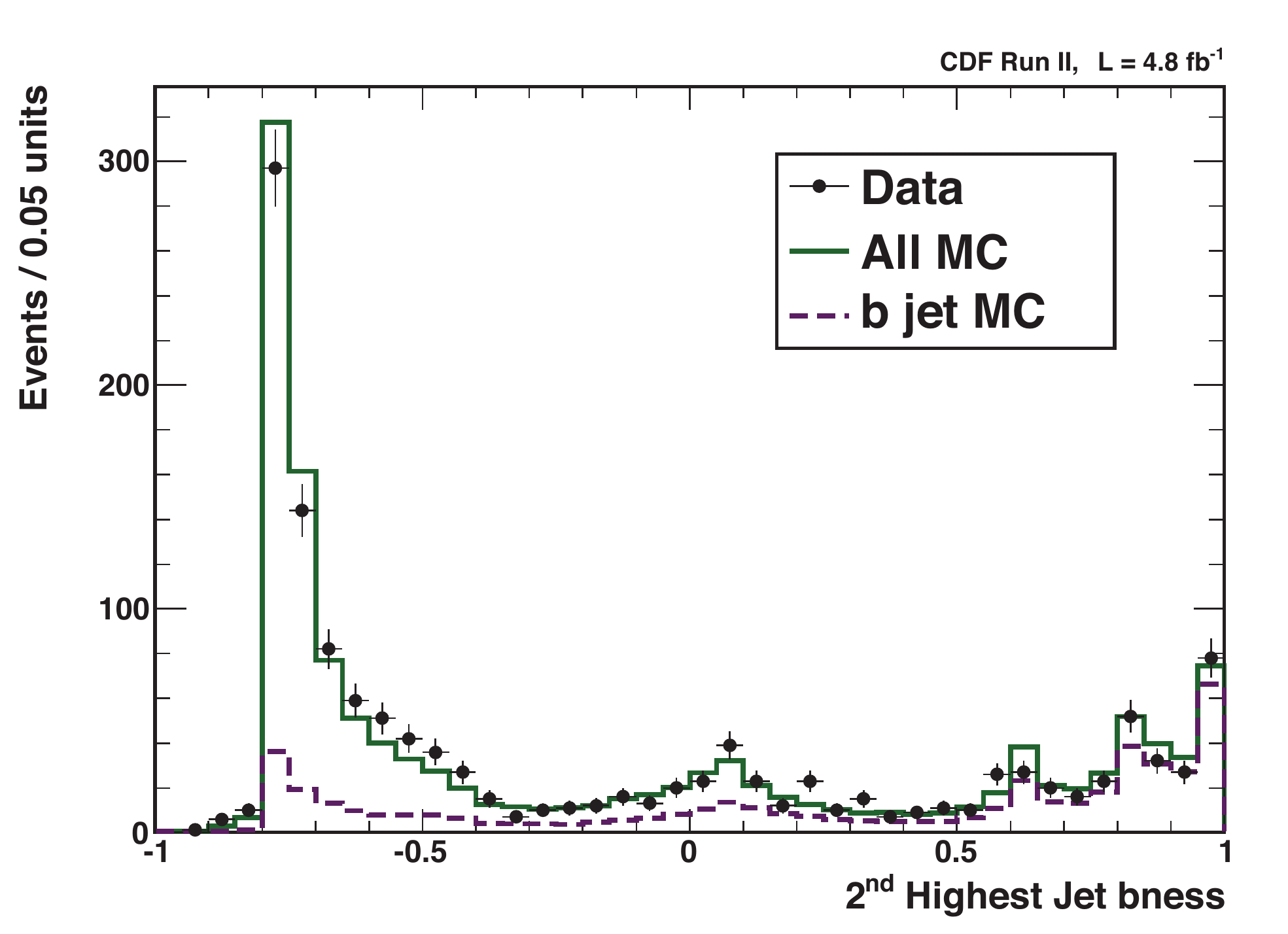}
  \includegraphics[width=0.48\textwidth]{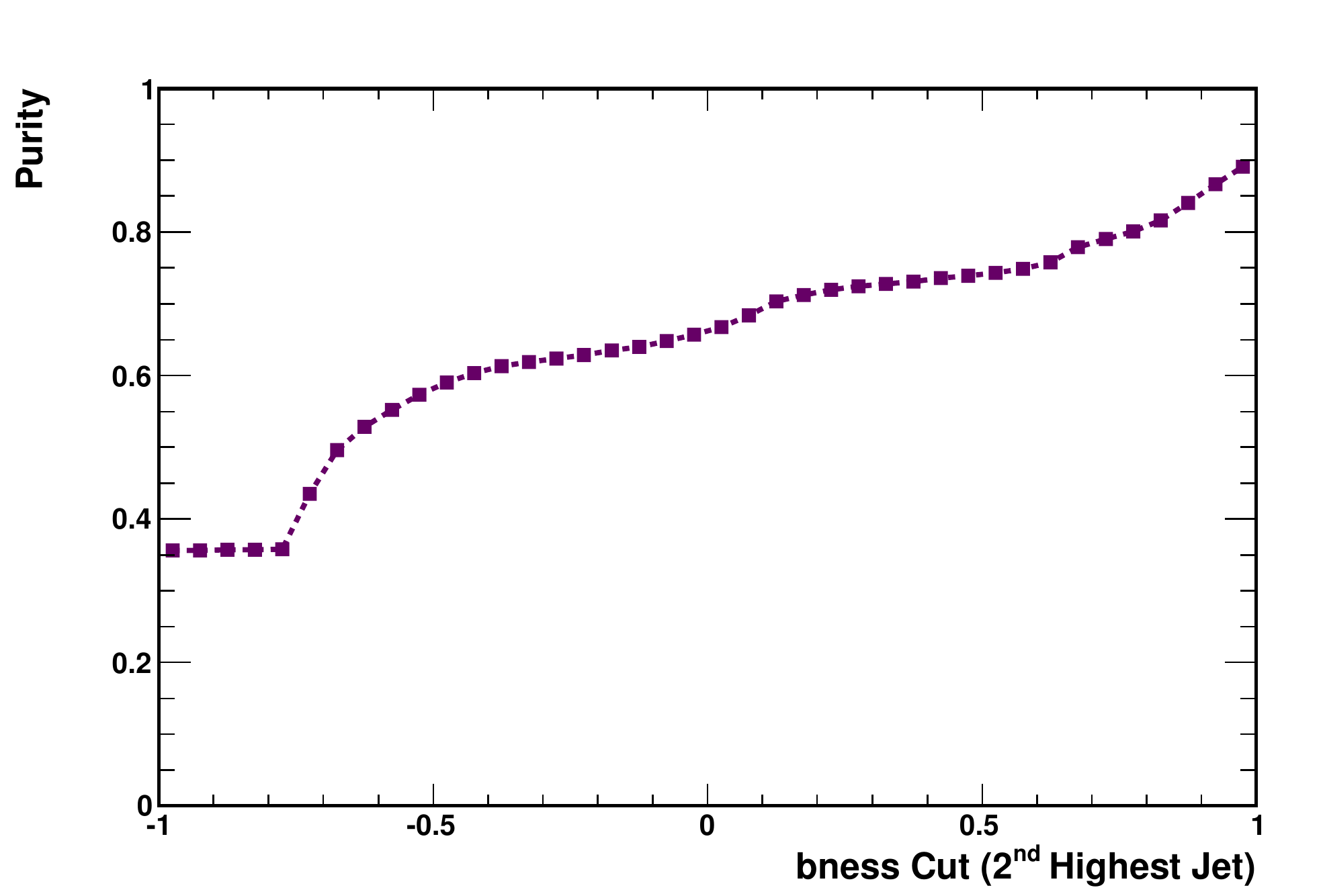}
  \caption{\textit{Top Left}: A comparison of the highest jet $b$ness
    in data (black points) and MC (green solid line) in the $t\bar{t}$
    lepton + jets sample, with the portion of the MC jets matched to
    $b$ quarks (purple dashed line) shown independently.  \textit{Top
      Right}: The $b$-jet purity for a given $b$ness cut on the
    highest jet $b$ness, as determined from matched jets in the MC.
    \textit{Bottom Left}: A comparison of the second highest jet
    $b$ness in data (black points) and MC (green solid line) in the
    $t\bar{t}$ lepton + jets selection region, with the portion of the
    MC jets matched to $b$ quarks (purple dashed line) shown
    independently.  \textit{Bottom Right}: The $b$-jet purity for a
    given $b$ness cut on the second highest jet $b$ness, as determined
    from matched jets in the MC.  In these plots, we see a high purity
    in our chosen sample, which is approximately 55\% $t\bar{t}$
    events.}
\label{fig:w2jet-jet_bness-Cut2-Matches}
\end{figure*}

We use our $t\bar{t}$ selection, described in
Section~\ref{sec:selection} and Table~\ref{tab:selection}, to
calculate the efficiency from a sample of jets with high $b$ purity. As these
events have many jets, we order the jets by decreasing
$b$ness value. This mirrors the procedure in a related
analysis using this $b$ tagger~\cite{metbb} and provides values
for the $b$-tagging efficiency while accounting for this sorting
procedure. Figure~\ref{fig:w2jet-jet_bness-Cut2} shows the jet
$b$ness distributions in data and MC for the two jets with highest $b$ness in each event.
The agreement here is very good, and regions of high
$b$ness are almost exclusively populated by $t\bar{t}$ events,
indicating that our $b$ tagger is properly identifying $b$ jets. We check that the
purity of $b$ jets as a function of the cut on the jet $b$ness in these
distributions is also high by splitting jets into matched and
non-matched categories
(Figure~\ref{fig:w2jet-jet_bness-Cut2-Matches}), as done for
the $Z + 1$ jet selection described in Section~\ref{sec:mistag}. We see
that the $b$-jet purity of the $t\bar{t}$ sample is rather high, even for low
$b$ness thresholds.

\begin{figure*}[tbp]
  \centering
  \includegraphics[width=0.48\textwidth]{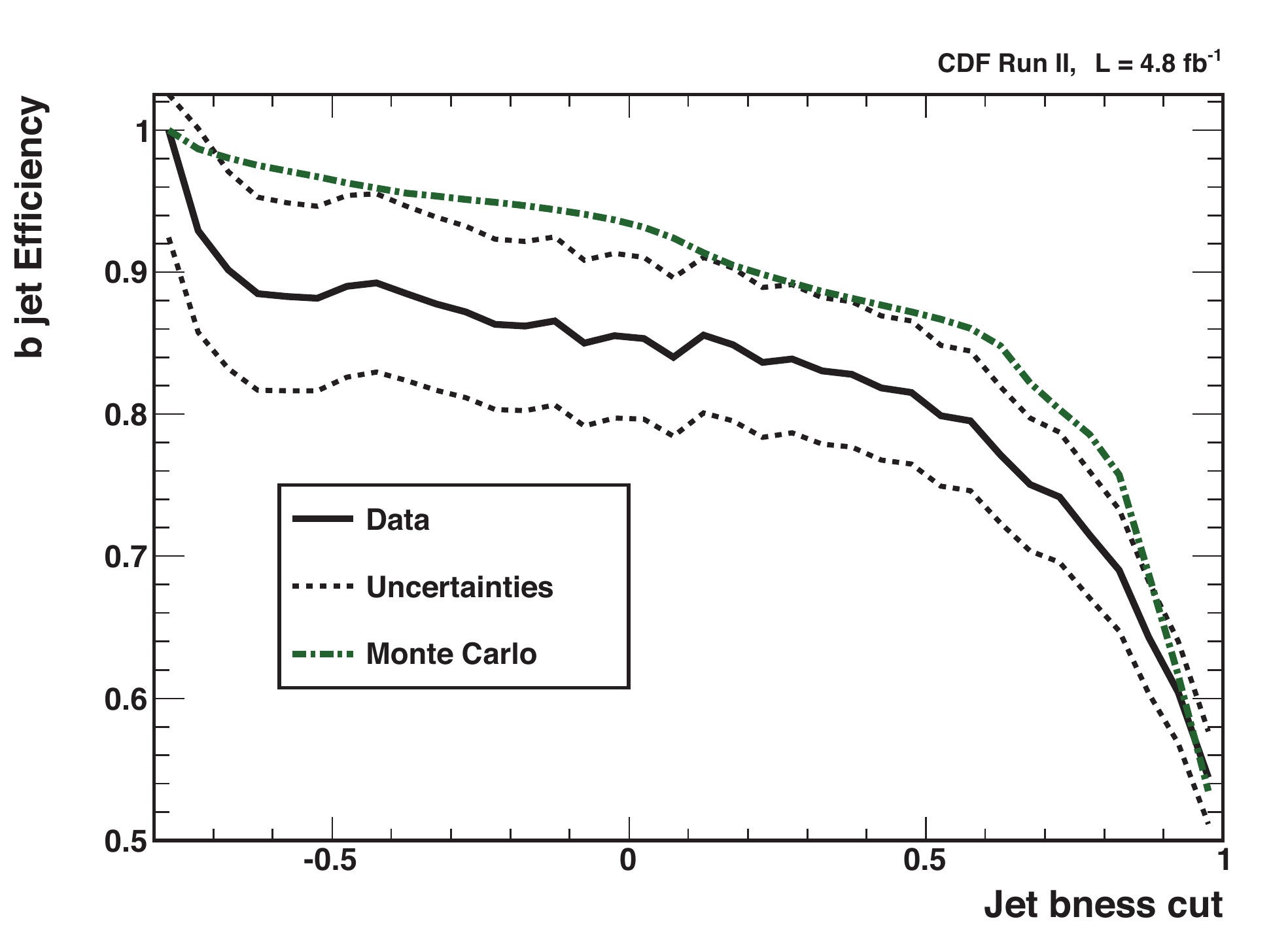}
  \includegraphics[width=0.48\textwidth]{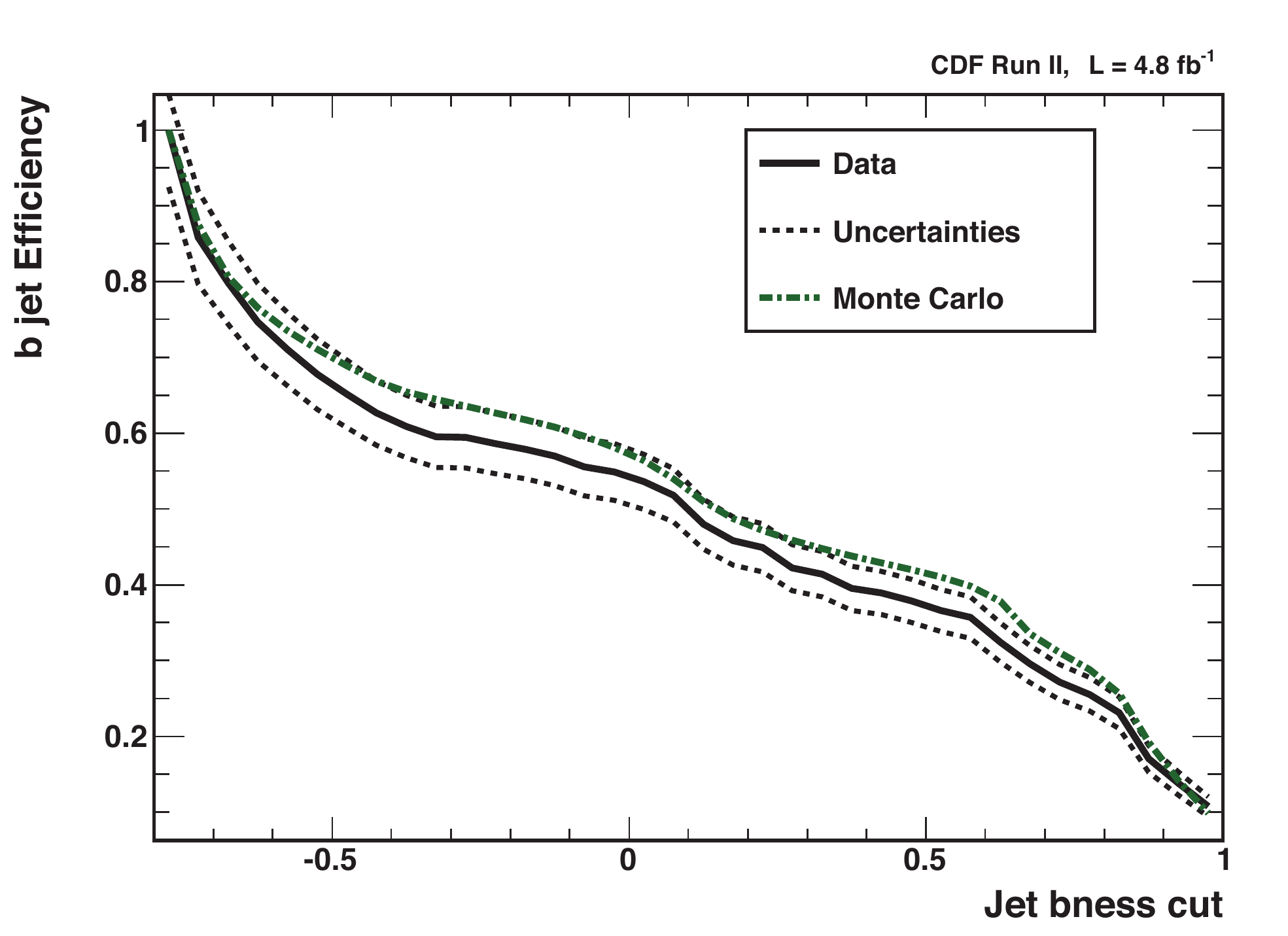}
\caption{The efficiency of a $b$ness cut in data (solid black line, dashed lines 
represent uncertainty) and Monte Carlo (dot-dashed green line) as a function of a cut 
on jet $b$ness for the highest (left) and 2\textsuperscript{nd} highest
(right) $b$ness jets in an event. We see our simulation typically over-predicts the efficiency measured in data, and thus needs to be corrected.}
\label{fig:jet_bness-efficiency}
\end{figure*}

\begin{figure*}[tbp]
  \centering
  \includegraphics[width=0.48\textwidth]{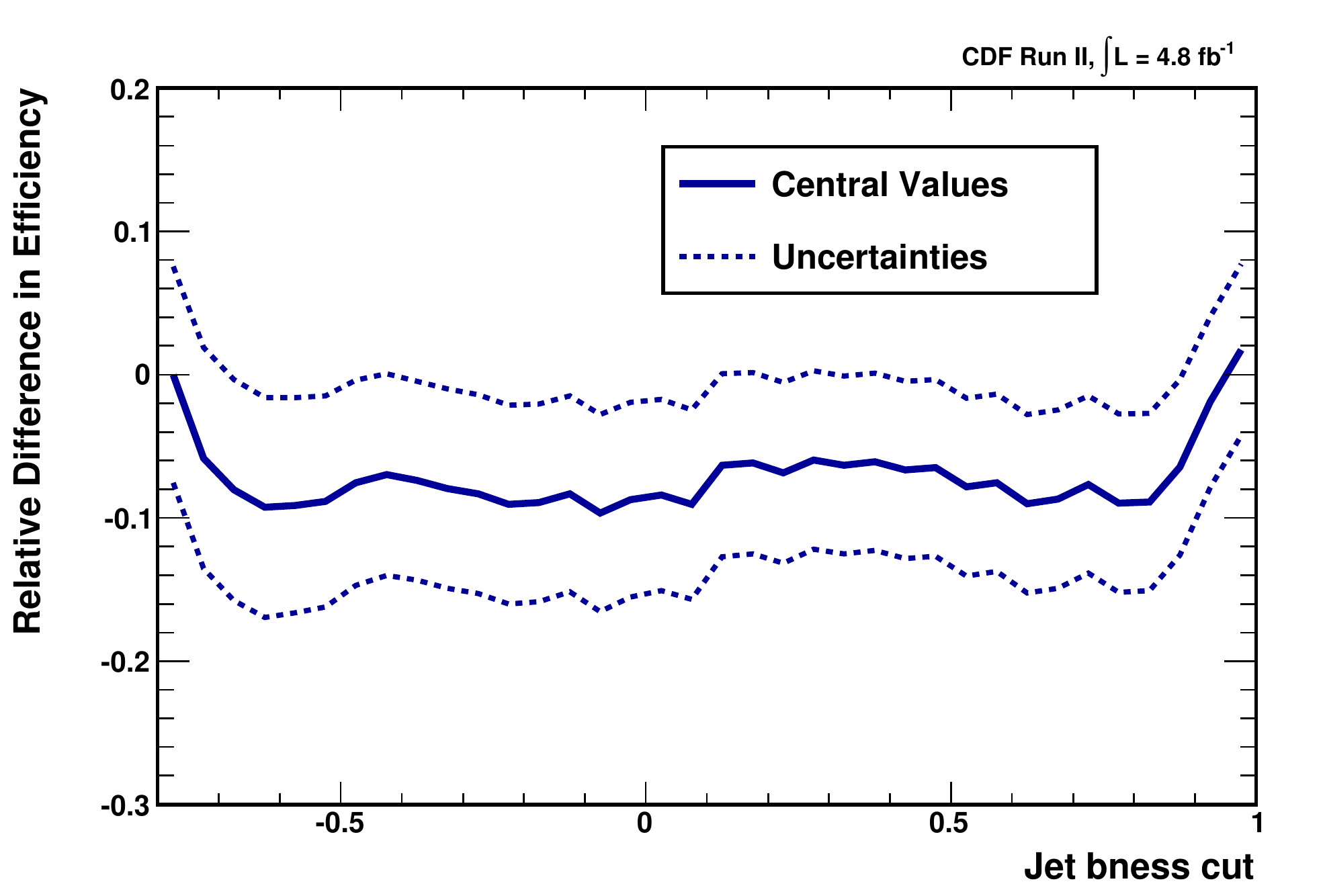}
  \includegraphics[width=0.48\textwidth]{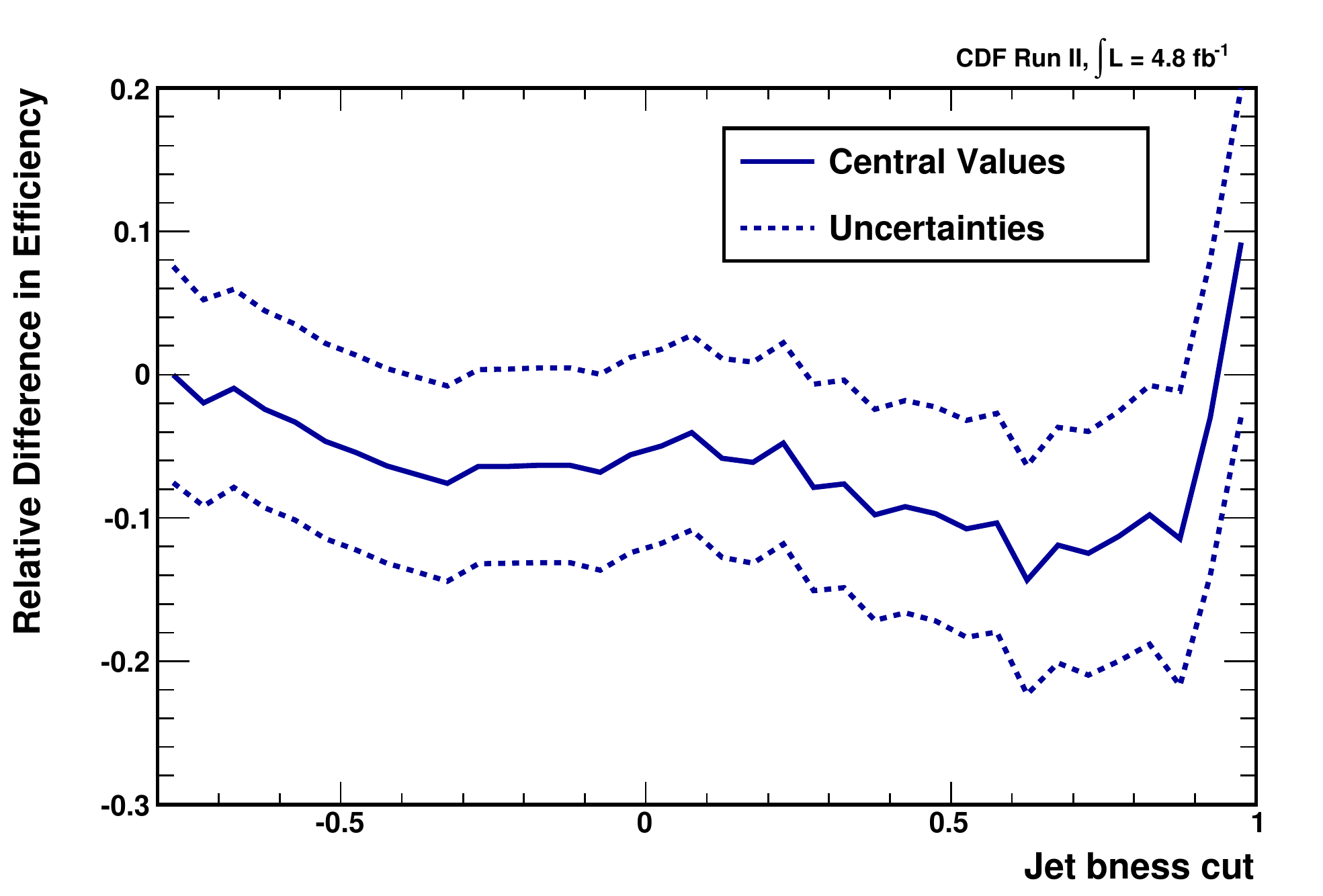}
\caption{The difference in efficiency between data and Monte Carlo (center solid 
line) and its uncertainty (dashed lines) relative to the efficiency in the Monte 
Carlo as a function of the cut on jet $b$ness for the highest (left) and 
2\textsuperscript{nd} highest (right) $b$ness jets in an event. The value of the 
scale factors and their uncertainties at the relevant $b$ness cuts in this 
analysis are summarized in Table \ref{tab:btag-values}.}
\label{fig:jet_bness-efficiencySF}
\end{figure*}

\begin{figure*}[tbp]
  \centering
  \includegraphics[width=0.48\textwidth]{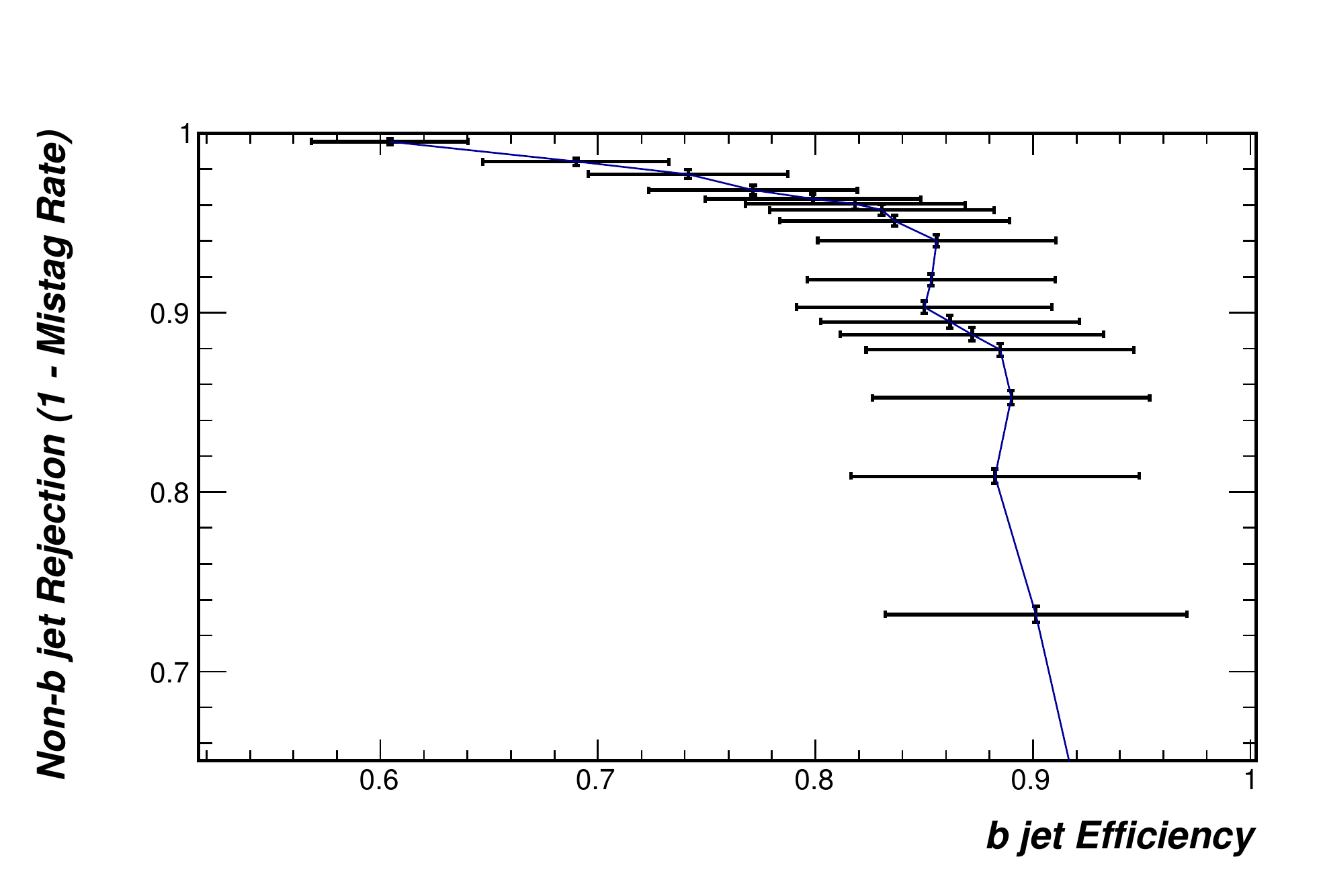}
  \includegraphics[width=0.48\textwidth]{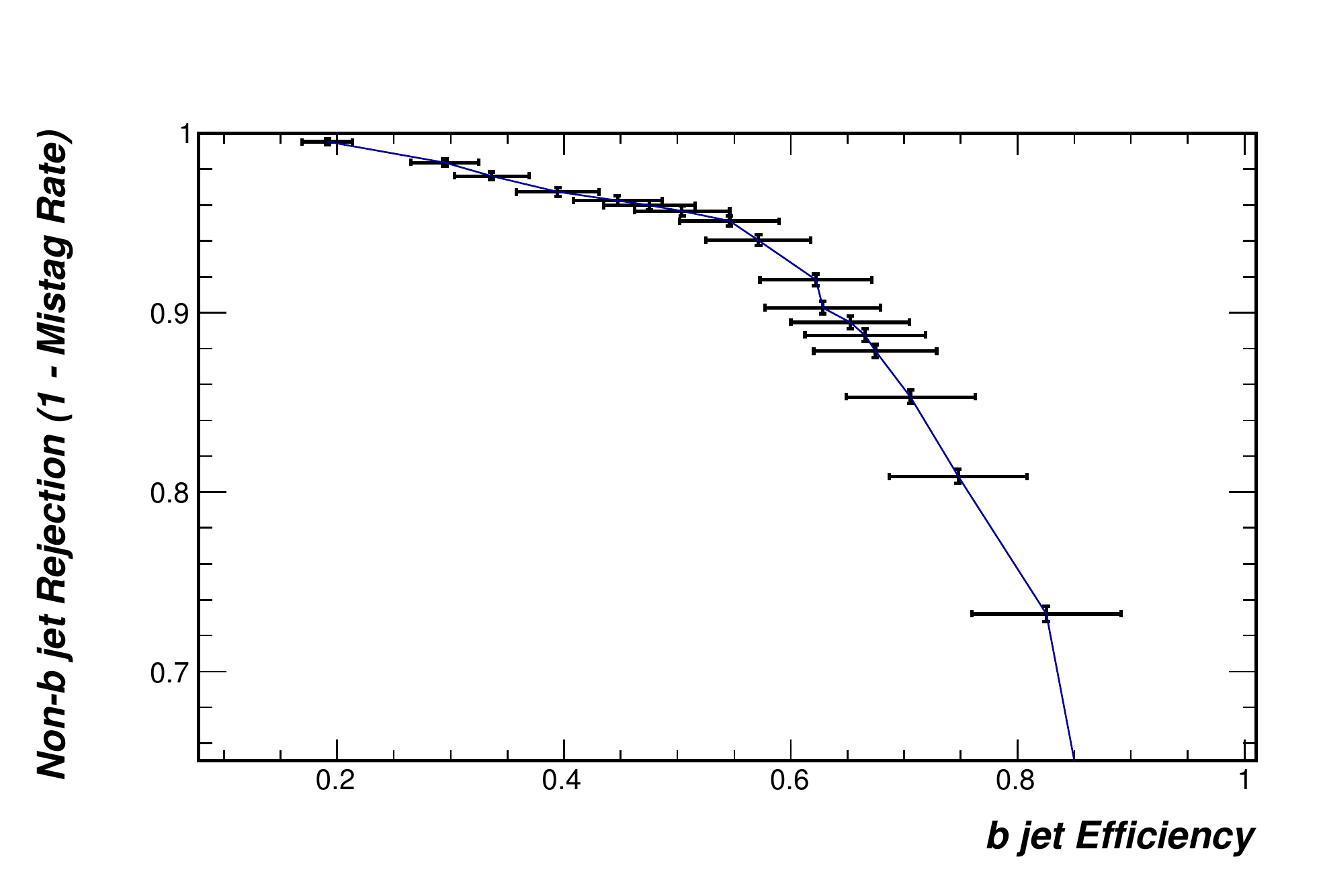}
\caption{Plots of the non-$b$-jet rejection versus the $b$-jet 
efficiency for a range of a cuts on jet $b$ness for the highest (left) and 
2\textsuperscript{nd} highest (right) $b$ness jets in an event.}
\label{fig:jet_bness-ROC}
\end{figure*}

We calculate the efficiency of a given $b$ness threshold and its
uncertainty in an analogous way to the calculation of the mistag rate,
described in detail in~\ref{sec:equations2}.  We show the calculated
efficiencies and uncertainties for the highest and
2\textsuperscript{nd} highest $b$ness jets in
Figure~\ref{fig:jet_bness-efficiency}, and we show the relative
difference between the efficiency in data and MC (the quantity
$s_{e}(b) -1$) and its uncertainty in
Figure~\ref{fig:jet_bness-efficiencySF}. The relative differences and
uncertainties on the efficiency are on the order of 10\% or less,
comparable to the SecVtx $b$ tagger scale factors and their
uncertainties.  Table~\ref{tab:btag-values} lists the efficiency and
mistag rates in data and MC for a chosen operating point---the highest
jet $b$ness $> 0.85$, and the 2\textsuperscript{nd} highest jet
$b$ness $> 0.0$---along with the relative difference between data and
MC, and the error on that difference. This choice of operating points
is motivated by the optimization of a cross section measurement that
uses the tagger~\cite{metbb}. Figure~\ref{fig:jet_bness-ROC} shows the
relationship between the calculated efficiency of identifying $b$ jets
with a cut on the jet $b$ness and the rejection power of that cut for
non-$b$ jets for the highest and 2\textsuperscript{nd} highest $b$ness
jets in an event.

\begin{table*}[hbt]
\begin{center}
\begin{tabular}{ll|llll} \hline
Quantity & $b$ness Cut & Data & MC & \% Difference & \% Error \\ \hline 
Mistag Rate & $0.0$ & $0.0819$ & $0.0720$ & $14 \%$ & $4.1 \%$ \\ 
& $0.85$ & $0.00997$ & $0.00869$ & $15 \%$ & $21 \%$ \\ \hline 
Tag Efficiency &  $0.0$ & $0.622$ & $0.684$ & $-9.0 \%$ & $8.7 \%$ \\ 
& $0.85$ & $0.652$ & $0.687$ & $-5.2 \%$ & $6.2 \%$ \\ \hline
\end{tabular}
\caption{Mistag rates and efficiencies on jet $b$ness cuts determined from comparisons of data and MC in our $Z + 1$ jet and $t\bar{t}$ control regions. For the $b$ness cut at 0.85, we consider the highest $b$ness jet, and for the $b$ness cut at 0.0, we consider the 2\textsuperscript{nd} highest $b$ness jet in our $t\bar{t}$ sample.}
\label{tab:btag-values}
\end{center}
\end{table*}

\begin{figure*}[tbp]
  \centering
  \includegraphics[width=0.49\textwidth]{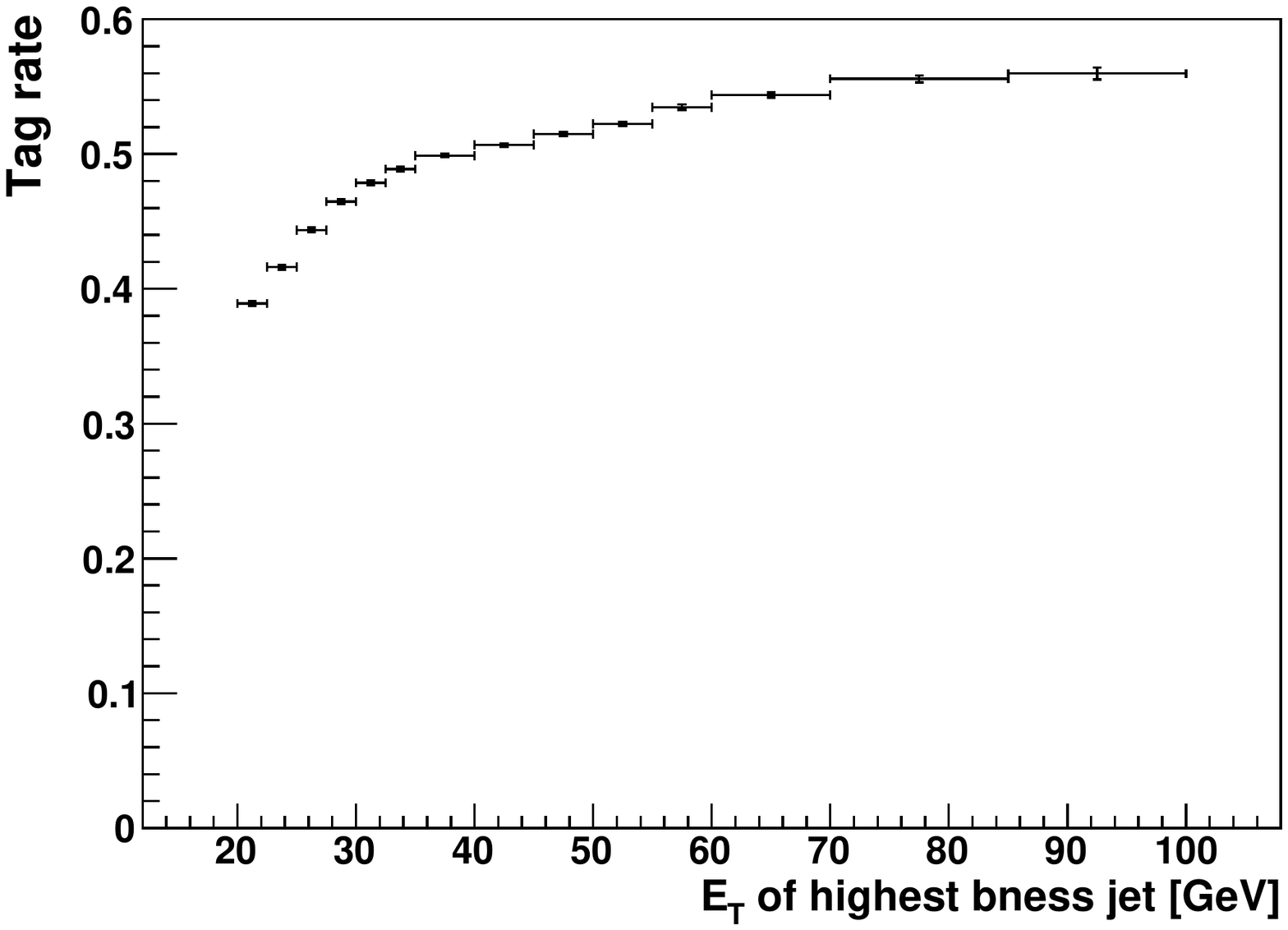}
  \includegraphics[width=0.49\textwidth]{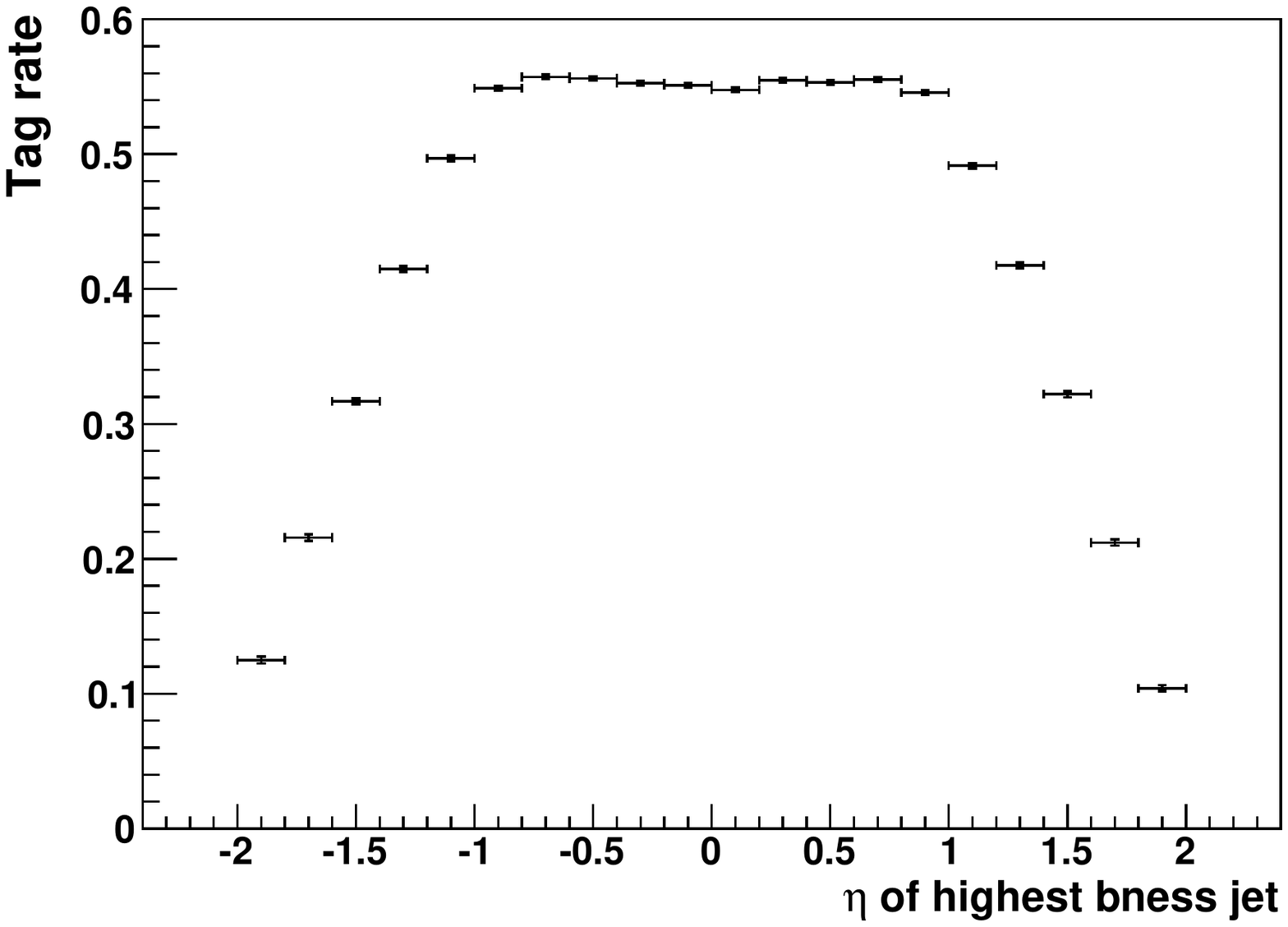}
  \caption{Tag performance for the jet with the highest $b$ness score
    as a function of transverse momentum (left) and $\eta$ (right) for
    a $b$ness requirement $b>0.85$, derived from simulated data.  The
    tagging efficiency ranges from 38\% at low transverse momentum to
    more than 50\% at higher momentum. The efficiency is flat in the
    central region ($|\eta|<1.0$) and drops outside the acceptance of
    the central part of the tracking system.  }
\label{fig:tag_performance}
\end{figure*}

We estimate the performance of the tagger as a function of the jets' transverse momenta
and pseudorapidity in simulated data of di-jet events, where the jets are $b$ jets.
We select the jet with the highest $b$ness
score and calculate the efficiency for $b$ness $>0.85$.  These efficiencies are shown in Figure~\ref{fig:tag_performance}.
The tagging efficiency ranges from 38\%
at low transverse momentum to more than 50\% at higher momentum. The
efficiency is flat in the central region ($|\eta|<1.0$) and drops
outside the acceptance of the central part of the tracking system.


While generic comparisons between taggers are difficult, we compare
our tagger to the most commonly used $b$ tagger in CDF, the SecVtx tagger.
The efficiency and mistag rates of our tagger compare favorably to the
SecVtx tagger.  We compare the two taggers using simulated events,
looking at the two highest $b$ness jets in the MC of our $t\bar{t}$
selection, and look at require $b>0.85$ for our tagger.  The ``tight"
SecVtx tagger operating point on this sample of jets has an efficiency
of $0.59$ and a mistag rate of $0.052$, while the ``loose" operating
point has an efficiency of $0.68$ with a mistag rate of $0.088$. For
the highest $b$ness jet cut at $>0.85$, we have a efficiency near the
loose-tag efficiency ($0.69$), but a lower mistag rate ($0.009$) than
the tight SecVtx tag; for the 2\textsuperscript{nd} highest $b$ness
jet cut at $>0.0$, we have a similarly high efficiency ($0.68$) while
allowing a mistag rate similar to the loose SecVtx tag ($0.082$).

\section{Conclusion}
\label{sec:conclusion}

We have described a neural network based $b$ tagger in current use at
the Fermilab Tevatron's CDF experiment. By examining all the tracks
associated with jets, this tagger has a larger acceptance than previous
neural network based taggers at CDF.
Furthermore, the tagger is calibrated using data from
$Z$ boson decays and events containing top quark pair production---a
novel method which yields small systematic uncertainties on the tagging
efficiency and mistag rate.
Finally, the utility of this tagger has been demonstrated in a
measurement of the $ZZ$ and $WZ$ production cross sections~\cite{metbb}.

\section*{Acknowledgements}
The authors thank the CDF collaboration, the Fermilab staff and the
technical staffs of the participating institutions for their vital
contributions. This work was supported by the US Department of Energy,
the US National Science Foundation and the Alfred P. Sloan Foundation.

\bibliographystyle{model1a-num-names}
\bibliography{prl}

\begin{thebibliography}{25}
\expandafter\ifx\csname natexlab\endcsname\relax\def\natexlab#1{#1}\fi
\providecommand{\bibinfo}[2]{#2}
\ifx\xfnm\relax \def\xfnm[#1]{\unskip,\space#1}\fi
\bibitem[{Abe et~al.(1994)}]{cdftop_evidence}
\bibinfo{author}{F.~Abe}, et~al., \bibinfo{journal}{Phys. Rev. D}
  \bibinfo{volume}{50} (\bibinfo{year}{1994}) \bibinfo{pages}{2966}.
\bibitem[{Abazov et~al.(2010)}]{d0tagging}
\bibinfo{author}{V.~M. Abazov}, et~al., \bibinfo{journal}{Nucl. Instrum.
  Methods A} \bibinfo{volume}{620} (\bibinfo{year}{2010}) \bibinfo{pages}{490}.
\bibitem[{{CMS Collaboration}(2011)}]{CMS-PAS-BTV-11-001}
\bibinfo{author}{{CMS Collaboration}}, \bibinfo{journal}{CMS Physics Analysis
  Summary}
  \bibinfo{volume}{\href{http://cdsweb.cern.ch/record/1366061}{CMS-PAS-BTV-11-001}}
  (\bibinfo{year}{2011}).
\bibitem[{{ATLAS Collaboration}(2011)}]{ATLAS-CONF-2011-102}
\bibinfo{author}{{ATLAS Collaboration}}, \bibinfo{journal}{ATLAS CONF Note}
  \bibinfo{volume}{\href{http://cdsweb.cern.ch/record/1369219}{ATLAS-CONF-2011-102}}
  (\bibinfo{year}{2011}).
\bibitem[{Acosta et~al.(2005)}]{secvtx}
\bibinfo{author}{D.~Acosta}, et~al., \bibinfo{journal}{Phys. Rev. D}
  \bibinfo{volume}{71} (\bibinfo{year}{2005}) \bibinfo{pages}{052003}.
\bibitem[{Abulencia et~al.(2006)}]{JetProb}
\bibinfo{author}{A.~Abulencia}, et~al., \bibinfo{journal}{Phys. Rev. D}
  \bibinfo{volume}{74} (\bibinfo{year}{2006}) \bibinfo{pages}{072006}.
\bibitem[{Acosta et~al.(2005)}]{slt}
\bibinfo{author}{D.~Acosta}, et~al., \bibinfo{journal}{Phys. Rev. D}
  \bibinfo{volume}{72} (\bibinfo{year}{2005}) \bibinfo{pages}{032002}.
\bibitem[{Richter(2007)}]{Richter:2007zzc}
\bibinfo{author}{S.~Richter}, \bibinfo{journal}{FERMILAB-THESIS-2007-35}
  (\bibinfo{year}{2007}).
\bibitem[{Aaltonen et~al.(2010)}]{Aaltonen:2010jr}
\bibinfo{author}{T.~Aaltonen}, et~al., \bibinfo{journal}{Phys. Rev. D}
  \bibinfo{volume}{82} (\bibinfo{year}{2010}) \bibinfo{pages}{112005}.
\bibitem[{Ferrazza(2006)}]{ferrazza06}
\bibinfo{author}{C.~Ferrazza}, \bibinfo{title}{{Identificazione di quark
  pesanti in getti adronici in interazioni \ppbar con il rivelatore CDF al
  Tevatron}}, Master's thesis, Universit\`{a} ``La Sapienza" Roma,
  \bibinfo{year}{2006}.
\bibitem[{Mastrandrea(2008)}]{Mastrandrea:2008zz}
\bibinfo{author}{P.~Mastrandrea}, \bibinfo{journal}{FERMILAB-THESIS-2008-63}
  (\bibinfo{year}{2008}).
\bibitem[{Abulencia et~al.(2007)}]{CDF_detect_A}
\bibinfo{author}{A.~Abulencia}, et~al., \bibinfo{journal}{J. Phys. G}
  \bibinfo{volume}{34} (\bibinfo{year}{2007}) \bibinfo{pages}{2457}.
\bibitem[{Affolder et~al.(2004)}]{cot_nim}
\bibinfo{author}{T.~Affolder}, et~al., \bibinfo{journal}{Nucl. Instrum. Methods
  A} \bibinfo{volume}{526} (\bibinfo{year}{2004}) \bibinfo{pages}{249}.
\bibitem[{Abe et~al.(1992)}]{jetclu}
\bibinfo{author}{F.~Abe}, et~al., \bibinfo{journal}{Phys. Rev. D}
  \bibinfo{volume}{45} (\bibinfo{year}{1992}) \bibinfo{pages}{1448}.
\bibitem[{Bhatti et~al.(2006)}]{jesnim}
\bibinfo{author}{A.~Bhatti}, et~al., \bibinfo{journal}{Nucl. Instrum. Methods
  A} \bibinfo{volume}{566} (\bibinfo{year}{2006}) \bibinfo{pages}{375}.
\bibitem[{Brun et~al.(1978)}]{geant3}
\bibinfo{author}{R.~Brun}, et~al., \bibinfo{title}{\textsc{geant3} manual},
  \bibinfo{year}{1978}. \bibinfo{note}{{CERN} Report CERN-DD-78-2-REV
  (unpublished)}.
\bibitem[{Campbell and Ellis(1999)}]{campbell}
\bibinfo{author}{J.~M. Campbell}, \bibinfo{author}{R.~K. Ellis},
  \bibinfo{journal}{Phys. Rev. D} \bibinfo{volume}{60} (\bibinfo{year}{1999})
  \bibinfo{pages}{113006}.
\bibitem[{Pumplin et~al.(2002)}]{cteq6m}
\bibinfo{author}{J.~Pumplin}, et~al., \bibinfo{journal}{J. High Energy Phys.}
  \bibinfo{volume}{0207} (\bibinfo{year}{2002}) \bibinfo{pages}{012}.
\bibitem[{Sj\"ostrand et~al.(2006)}]{pythia}
\bibinfo{author}{T.~Sj\"ostrand}, et~al., \bibinfo{journal}{J. High Energy
  Phys.} \bibinfo{volume}{05} (\bibinfo{year}{2006}) \bibinfo{pages}{026}.
\bibitem[{Hoecker et~al.(2007)Hoecker, Speckmayer, Stelzer, Therhaag, von
  Toerne, and Voss}]{tmva}
\bibinfo{author}{A.~Hoecker}, \bibinfo{author}{P.~Speckmayer},
  \bibinfo{author}{J.~Stelzer}, \bibinfo{author}{J.~Therhaag},
  \bibinfo{author}{E.~von Toerne}, \bibinfo{author}{H.~Voss},
  \bibinfo{journal}{PoS} \bibinfo{volume}{ACAT} (\bibinfo{year}{2007})
  \bibinfo{pages}{040}.
\bibitem[{Aaltonen et~al.(2009)}]{metjj_prl}
\bibinfo{author}{T.~Aaltonen}, et~al., \bibinfo{journal}{Phys. Rev. Lett.}
  \bibinfo{volume}{103} (\bibinfo{year}{2009}) \bibinfo{pages}{091803}.
\bibitem[{Goncharov et~al.(2006)}]{emtiming}
\bibinfo{author}{M.~Goncharov}, et~al., \bibinfo{journal}{Nucl. Instrum.
  Methods A} \bibinfo{volume}{565} (\bibinfo{year}{2006}) \bibinfo{pages}{543}.
\bibitem[{Mangano et~al.(2003)}]{alpgen}
\bibinfo{author}{M.~L. Mangano}, et~al., \bibinfo{journal}{J. High Energy
  Phys.} \bibinfo{volume}{07} (\bibinfo{year}{2003}) \bibinfo{pages}{001}.
\bibitem[{Aaltonen et~al.(2009)}]{ZbPRD}
\bibinfo{author}{T.~Aaltonen}, et~al., \bibinfo{journal}{Phys. Rev. D}
  \bibinfo{volume}{79} (\bibinfo{year}{2009}) \bibinfo{pages}{052008}.
\bibitem[{Aaltonen et~al.(2011)}]{metbb}
\bibinfo{author}{T.~Aaltonen}, et~al., \bibinfo{journal}{arXiv:1108.2060}
  (\bibinfo{year}{2011}).

\end{thebibliography}

\appendix
\section{Evaluation of Mistag Rate and Efficiency}
\label{sec:equations}

For any given selection of data, we can
calculate the mistag rate (where all non-$b$ jets are considered
mistags) if we know the number $N_{B}$ of $b$ jets, the number  $N_{B}(b)$ of $b$
jets above the threshold $b$ness, the total number $N$ of
jets, and the total number $N(b)$ of jets above the $b$ness cut
threshold:
\begin{equation}
m(b)=\frac{N(b) - N_{B}(b)}{N - N_{B}}.
\label{eq:mistag}
\end{equation}
We may use MC to determine the fraction $f_{B}$ of jets that are $b$
jets, and the efficiency $e_\text{MC}(b)$ for these jets to pass the $b$ness cut. 
This efficiency may need to be modified by a scale
factor $s_{e}(b) = e(b) / e_\text{MC}(b)$ if it is different from the true
efficiency evaluated in data. Thus,
\begin{equation}
N_{B}=f_{B}N  \text{~~and~~}  N_{B}(b) = s_{e}(b)e_\text{MC}(b)f_{B}N.
\end{equation}
Also, if we define a mistag rate that has not been corrected for the possible 
presence of $b$ jets in the same sample, $m_\text{raw}(b) = N(b)/N$, then we may write 
equation \ref{eq:mistag} in the following way:
\begin{align}
m(b) &= \frac{m_\text{raw}(b)N - s_{e}(b)e_\text{MC}(b)f_{B}N}{N - f_{B}N} \notag \\
     &= \frac{m_\text{raw}(b) - s_{e}(b)e_\text{MC}(b)f_{B}}{1 - f_{B}}.
\label{eq:mistag_final}
\end{align}
We can write an analogous expression for the efficiency of $b$ jets passing a 
given $b$ness cut:
\begin{equation}
e(b) = \frac{e_\text{raw}(b) - s_{m}(b)m_\text{MC}(b)f_{L}}{1 - f_{L}}
\label{eq:efficiency}
\end{equation}
where $e_\text{raw}(b)$ is a ``raw" efficiency uncorrected for the presence of non-$b$ 
jets in a sample, $m_\text{MC}(b)$ is the mistag rate as measured in MC, corrected to 
match data by a scale factor $s_{m}(b)$, and $f_{L}$ is the fraction of
light-flavor (here defined as non-$b$) jets in the chosen sample. 

Note that the determination of the mistag rate depends on the calculated value of 
the efficiency (through the scale factor term $s_{e}(b)$), and that in turn the 
determination of the efficiency depends on the mistag rate (again through the 
scale factor $s_{m}(b)$). Similarly, the uncertainties on these quantities (see below) 
depend on each other in a non-linear fashion. Thus, we use an iterative procedure 
to solve for the mistag rate, efficiency, and their uncertainties. 
We calculate the mistag 
rate first using a value of $s_{e}(b)=1$, and find that the values of $e(b)$ and 
$m(b)$ converge (and their uncertainties) very quickly.

The uncertainties on these quantities may also be calculated from the
expressions above. For the mistag rate, 
\begin{align}
    \sigma_{m}^{2}(b) &= \frac{m_\text{raw}(b)(1-m_\text{raw}(b))}{N(1-f_{B})^{2}} \notag \\
    &+ \left(\frac{\sigma_{e}(b)f_{B}}{1-f_{B}}\right)^{2} \notag \\
    &+ \left(\frac{\sigma_{f_{B}}[s_{e}(b)e(b)- m(b)]}{1-f_{B}}\right)^{2}.
    \label{eq:mistag_uncertainty}
\end{align}
The first term is a binomial uncertainty on the raw mistag rate of the sample, 
and is the term related to the statistical uncertainty of the sample used to 
determine the mistag rate. The second term comes from the uncertainty on the 
measured value of $e(b)$, which can be calculated using a similar expression, 
and is done so iteratively, as $\sigma_{m}(b)$ and $\sigma_{e}(b)$ depend on 
each other. The final term is due to the uncertainty on $f_{B}$, which will 
depend on the choice of MC and the region in which MC and data are compared. A 
similar expression determines $\sigma_{e}(b)$.

\section{Tagging Efficiency Determination}
\label{sec:equations2}

Similar to our calculation of the mistag rate, we calculate the
efficiency observed in data using equation \ref{eq:efficiency}. Both
$e_\text{raw}(b)$ and $s_{m}(b)m_\text{MC}(b)=m(b)$ can be calculated easily by
counting events above a given $b$ness threshold in the data and MC
respectively. Because of the different competing processes in our
$t\bar{t}$ sample (there is a significant contribution from $W$ + light
flavor jets and $W$ + $b\bar{b}$ processes), it is best to break $f_{L}$
into these most significant subsamples:
\begin{equation}
f_{L} = \frac{f_{L}^{Wjj}N_{Wjj} + f_{L}^{Wb\bar{b}}N_{Wb\bar{b}} + f_{L}^{t\bar{t}}N_{t\bar{t}}}{N_{Wjj} + N_{Wb\bar{b}} + N_{t\bar{t}} }
\label{eq:efficiency-fL}
\end{equation}
where $N_{X}$ is the number of events predicted by MC in subsample $X$, and 
$f_{L}^{X}$ is the fraction of non-$b$ jets in subsample $X$. We assume that the 
MC correctly reproduces the values of $f_{L}^{X}$. To determine $s_{e}(b) = e(b)/
e_\text{MC}(b)$, we write down a similar expression for the efficiency in MC using the 
efficiency of each subsample in MC:
\begin{equation}
e_\text{MC}(b) = \frac{1}{N_{Wjj} + N_{Wb\bar{b}} + N_{t\bar{t}}}\sum_{X} e_{X}(b)f_{B}^{X}N_{X}
\label{eq:efficiency-eMC}
\end{equation}
where, as before, $N_{X}$ is the number of events predicted by Monte Carlo in 
subsample $X$, $f_{B}^{X}$ is the total fraction of $b$ jets in subsample $X$, 
and $e_{X}$ is the efficiency of $b$ jets passing a particular $b$ness cut in 
subsample $X$. We assume, again, that the Monte Carlo correctly reproduces the 
values of $f_{B}^{X}$.

Given Equations \ref{eq:efficiency-fL} and \ref{eq:efficiency-eMC}, we modify our 
equation for determining the uncertainty in the calculated efficiency. We obtain the uncertainty 
by calculating the uncertainty of the quantity $(e(b) - e_\text{MC}(b))$, and find 
\begin{align}
  \sigma_{e}^{2}(b) &=
  \frac{1}{(1-f_{L})^{2}} \left(\frac{e_\text{raw}(1-e_\text{raw})}{N_{D}} 
    + (\sigma_{m}f_{L})^{2}\right) \notag \\
  &+ \sum_{X}\frac{\sigma_{X}^{2}}{\left[N_\text{MC}(1-f_{L})\right]^2} \times \notag \\
  &\left[(e+s_{m}m)(f_{L}-f_{L}^{X})\right. 
   \left.+f_{B}^{X}(e_\text{MC}-e_{X})\right]^{2} 
\label{eq:efficiency-unc}
\end{align}
where the latter term represents a sum over each of the MC subsamples. $N_\text{MC}$ 
and $N_{B}$ are the total number of events and events with $b$ jets in the MC, 
and $\sigma_{X}$ is the uncertainty assigned to the number of events in each MC 
subsample. Because we compare only the normalizations of data and MC in our 
determination of efficiency (and mistag rate) scale factors, the uncertainty on 
the number of events in each MC subsample need only reflect the relative 
uncertainty on the fraction of events each subsample contributes to the whole. We 
assign $\sigma_{Wb\bar{b}}=20\%$, and $\sigma_{Wjj}=8.72\%$ and $\sigma_{t
\bar{t}}=6.78\%$ based on a fit to the distribution of the sum of the highest two 
$b$ness jets in $t\bar{t}$ events.

\end{document}